\documentclass[reprint,twocolumn,footinbib,longbibliography,superscriptaddress]{revtex4-2}
\usepackage{amsmath}
\usepackage{amsfonts}
\usepackage{amssymb}
\usepackage{lmodern}
\usepackage{times}
\usepackage[pdftex]{graphicx}
\usepackage{bm}
\usepackage{amsthm}
\usepackage{mathtools}
\usepackage{physics}
\usepackage{mathrsfs}
\usepackage{hyperref}

\newcommand{\Sext}{S_\mathrm{ext}}

\newcommand{\x}{\textbf{x}}
\newcommand{\y}{\textbf{y}}

\newcommand{\Tst}{T_\mathrm{stv}}
\newcommand{\Tlag}{T_\mathrm{lag}}

\begin{document}

\title{
Universal Transitions between Growth and Dormancy via Intermediate Complex Formation
}

\author{Jumpei F. Yamagishi}
\affiliation{Department of Basic Science, The University of Tokyo, 3-8-1 Komaba, Meguro-ku, Tokyo 153-8902, Japan}
\author{Kunihiko Kaneko}
\affiliation{Center for Complex Systems Biology, Universal Biology Institute, University of Tokyo, 3-8-1 Komaba, Meguro-ku, Tokyo 153-0041, Japan} 
\affiliation{Niels Bohr Institute, University of Copenhagen, Blegdamsvej 17, 2100 Copenhagen, Denmark}

\begin{abstract}
    A simple cell model consisting of a catalytic reaction network with intermediate complex formation is numerically studied. As nutrients are depleted, the transition from the exponential growth phase to the growth-arrested dormant phase occurs along with hysteresis and a lag time for growth recovery. This transition is caused by the accumulation of intermediate complexes, leading to the jamming of reactions and the diversification of components. These properties are generic in random reaction networks, as supported by dynamical systems analyses of corresponding mean-field models.
\end{abstract}

\maketitle
\section{Introduction}
As microbial cells proliferate, they are crowded and nutrients in the environment are depleted. The cells then enter the dormant phase (or so-called stationary phase), in which cell growth is significantly arrested~\footnote{Such growth-arrested states are often termed dormant, non-growing, or quiescent phases/states, as well as stationary phase for an ensemble of cells. Herein, we adopt the term dormant phase}. This behavior is commonly observed across microbial species and even mammalian cells under a variety of environmental conditions~\cite{gray2004sleeping}. 
In fact, most microbial cells in natural ecosystems are in the growth-arrested dormant phase as they are under resource limitation~\cite{finkel2006long,navarro2010stationary,del2008physiological,gefen2014direct,kirchman2018processes}. 
Once cells enter the dormant phase, the intracellular metabolic phenotypes drastically change, and hysteresis and bistability between phenotypes with exponential and arrested growth are observed~\cite{kotte2014phenotypic,veening2008bistability,krishna2018minimal}; in addition, a certain time is required to recover growth even after the resource supply has resumed, and it is known as the lag time~\cite{levin2010automated,joers2016growth,kaplan2021observation}.

Despite the importance of such universal and mundane behavior, the theoretical understanding of dormancy is still in its infancy compared with that of the exponential growth phase, for which well-established quantitative theories are available~\cite{scott2011bacterial,kaneko2015universal,jun2018fundamental}. 
Although specific molecular mechanisms of dormancy have been extensively studied~\cite{navarro2010stationary,jaishankar2017molecular}, little attention has been paid to establishing a theory for universal characteristics of the dormant phase and transitions to it. 
Refs.~\cite{himeoka2017theory}~and~\cite{reich2022slow} represent a few early exceptions. 
In Ref.~\cite{himeoka2017theory}, by assuming that nutrient limitation leads to the accumulation of waste chemicals, a phenomenological model for the growth-dormant transition was proposed and quantitative laws of the lag time were derived. In Ref.~\cite{reich2022slow}, an abstract spin glass model for aging dynamics was proposed. 
However, the universality of their results and the origin(s) of these specific mechanisms have not been fully explored. 
Therefore, a better understanding of the growth-dormant transition as a universal behavior of cells growing through intracellular reactions with many components is required. 

In this Letter, by considering a simple cell model consisting of catalytic reactions of many components, we demonstrate that such a transition between growth and dormant phases generally appears without specifically tuning the intracellular reactions as long as intermediate complexes between substrates and catalysts have sufficient lifetimes. 
The transition is {aused by the accumulation of complexes under the depletion of nutrients, and it is characterized as a cusp bifurcation in dynamical systems theory. 
The transition observed in random reaction networks is then analyzed using the ``mean-field'' theory of catalytic reaction dynamics, which also implies that the transition to a dormant phase does not require any special mechanism and is a universal feature of cells that grow by intracellular catalytic reactions.

\section{Model}
In this study, we adopt a simple model of cellular dynamics that captures only the basic features of these dynamics. It consists of intracellular reaction networks and transport reactions of externally supplied nutrient(s). 
Complicated intracellular metabolic reactions are simplified as randomly connected catalytic reaction networks.
Although such models with catalytic reaction networks have reproduced the statistics of cells in the exponential growth phase~\cite{furusawa2003zipf,furusawa2012adaptation}, they do not demonstrate the growth-dormant transition. 
One possible drawback in these models is that catalytic reactions progress immediately. In reality, each chemical reaction progresses after the formation of an intermediate complex between the substrate and catalyst is formed. 

Here we introduce a model that includes the formation of intermediate complexes in reactions and examine whether and how the growth-dormant transition is exhibited by the model. Then, each catalytic reaction $\rho$, in which substrate $X_{\rho_s}$ is converted into product $X_{\rho_p}$ by catalyst $X_{\rho_c}$, consists of two-step elementary reaction processes with the formation of an intermediate complex $Y_\rho$ as follows:
\begin{eqnarray*} \label{eq:MM}
X_{\rho_s}+ X_{\rho_c} \rightleftharpoons^{k^+_\rho}_{k^-_\rho} 
Y_\rho 
\overset{v_\rho}{\to} X_{\rho_p}+ X_{\rho_c}. 
\end{eqnarray*}
Here, each elementary process proceeds according to the law of mass action with the labeled coefficient, and $\rho_s$, $\rho_p$, and $\rho_c$ denote the indices of the substrate, product, and catalyst for reaction $\rho$, respectively. 
For the adiabatic limit $v_\rho\to\infty$, the above reaction processes are reduced to the single mass action kinetics without intermediate complex formation, $X_{\rho_s}+ X_{\rho_c} \to X_{\rho_p} + X_{\rho_c}$, and the model is reduced to those studied earlier~\cite{furusawa2003zipf,furusawa2012adaptation}. 
In contrast, when $v_\rho$ is small, the intermediate complex $Y_\rho$ can accumulate, leading to a decrease in free reactants that are not bound into complexes, which can hinder the reaction processes. 

Considering a cell consisting of $n$ chemicals and $N_r$ reactions (and corresponding intermediate complexes), its state is represented by a set of concentrations $(\x,\y)$ of free reactants (that are not bound into complexes) $X_i$ and complexes $Y_\rho$. 
The time change of the cellular state $(\x,\y)$ is then given as follows: 
\begin{eqnarray}
\dot{x}_i &=&
\sum_{\rho}
(\delta_{i,\rho_p}+\delta_{i,\rho_c})v_\rho y_{\rho}-(\delta_{i,\rho_s}+\delta_{i,\rho_c})f_\rho({\bf x},{\bf y})
\nonumber\\
&& + F_i(\x;\Sext,\alpha) - \mu x_i, 
\label{eq:dxdt} \\
\dot{y}_\rho &=& f_\rho({\bf x},{\bf y}) - v_\rho y_{\rho} - \mu y_\rho, \label{eq:dydt}
\end{eqnarray}
where $f_\rho({\bf x},{\bf y}) := k^+_\rho x_{\rho_s}x_{\rho_c} - k^-_\rho y_{\rho}$ is the total consumption rate of substrate $\rho_s$ by reaction $\rho$, and $\delta$ is Kronecker's delta. The term $F_i(\x;\Sext,\alpha)$ in Eq.~(\ref{eq:dxdt}) represents the intake of chemical $X_i$ ($i=0,1,\cdots,n-1$), which can be non-zero if $X_i$ is a nutrient but is zero otherwise. The last terms in Eqs.~(\ref{eq:dxdt}-\ref{eq:dydt}), $-\mu x_i$ and $-\mu y_\rho$, represent the dilution of each concentration due to cellular volume growth. 
The growth rate $\mu$ is given by $\mu(\x,\y):=\sum_i F_i({\bf x})$, because for simplicity, we assumed that the contribution of each chemical $X_i$ to volume or weight is uniform regardless of $i$. $\sum_i x_i + 2\sum_{\rho}y_\rho = 1$ is then constant in the dynamics~(\ref{eq:dxdt}-\ref{eq:dydt}) based on the law of mass conservation. 

Below, for simplification purposes, the reaction rate constants $k^+_\rho$, $k^-_\rho$, and $v_\rho$ are set as independent of $\rho$, and they are denoted by $k^+=1$, $k^-=0$, and $v$, respectively. For simplicity, we also assumed that 
there is only a single nutrient chemical $X_0$. Its intake is mediated by transporter chemical $X_1$ with $\alpha=2$, i.e., $F_i(\x;\Sext,\alpha) = \delta_{i0}\Sext x_{1}^\alpha$, where $\Sext$ denotes the environmental concentration of nutrient chemical $X_0$, and the transport coefficient for $F_i$ is normalized as unity. 
Note that the following results and arguments hold independent of the details of settings, such as parameter values and specific functional forms of nutrient intake $F_i$ (see also Supplemental Material (SM), Sec.~\ref{sec:random_net}).

\begin{figure}[tbh]
    \centering 
    \includegraphics[width=\linewidth, clip]{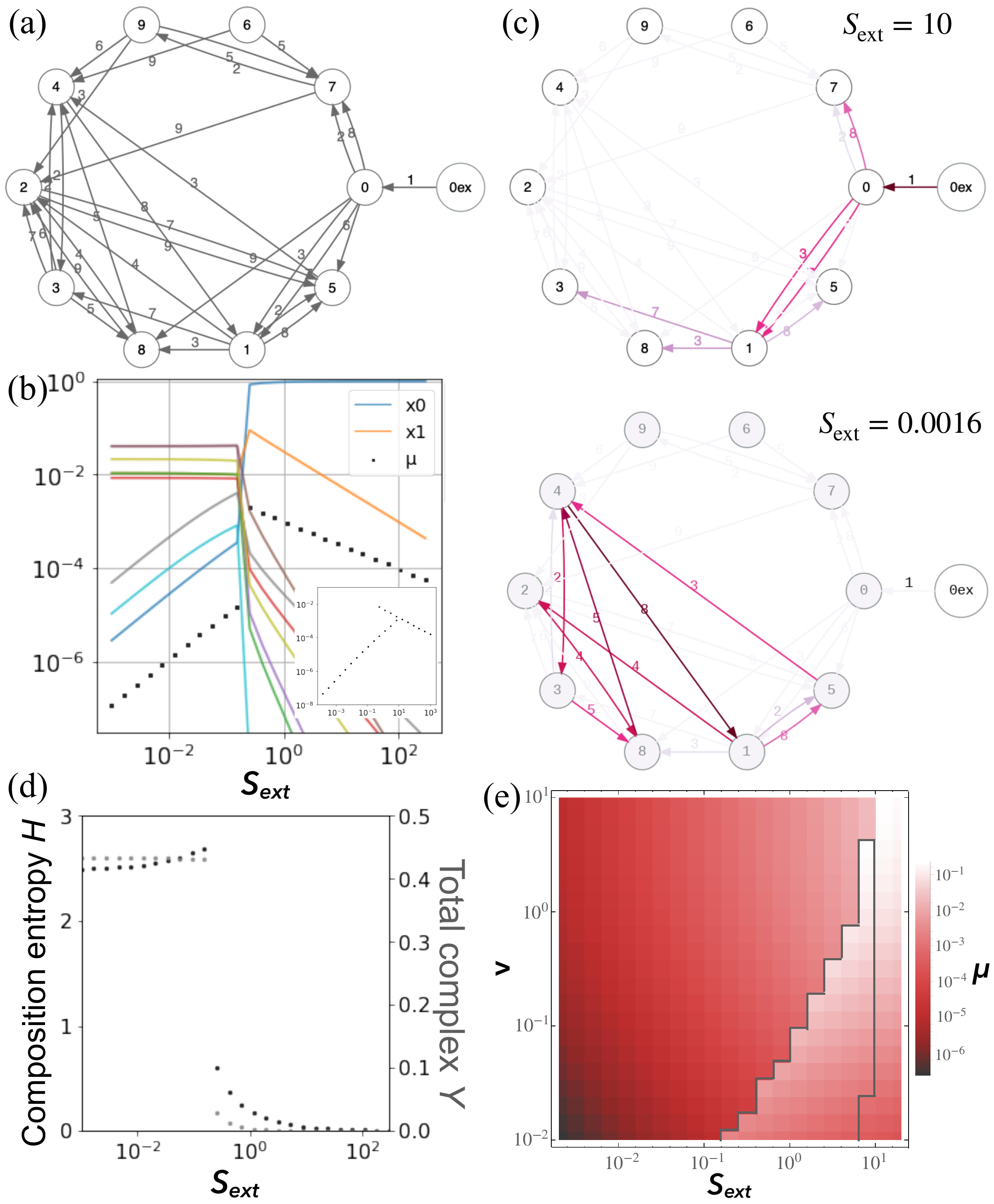}
    \caption{
    Example of a growth-dormant transition in randomly generated networks. (a) Reaction network ($n=10,N_r=30$). Chemicals at arrowtails are transformed into those at arrowheads, catalyzed by the chemicals labeled on the edges. 
	Nutrient $X_0$ is taken up via active transport by transporter $X_1$ in proportion to $x_1^2$. 
    (b) Dependence of $\mu^\ast$ (black points) and $\x^\ast$ (colored lines) on $\Sext$. $v=0.01$. 
	(inset) Hysteresis and bistability for $\mu^\ast$. $v=0.03$. 
    (c) Dominant pathways for the growth phase ($\Sext=10$; top) and dormant phase ($\Sext=0.0016$; bottom). $v=0.01$. 
	The edge colors represent reaction fluxes in the log scale. 
    (d) Dependence of composition entropy $H := -\sum_ix_i\log x_i-\sum_\rho2y_\rho\log(2y_\rho)$ (black) and $Y := \sum_\rho y_\rho$ (gray) on $\Sext$. $v=0.01$. 
    (e) Dependence of $\mu^\ast$ on $(\Sext,v)$. 
    $\mu^\ast$ is numerically calculated by decreasing $\Sext$ for each $v$, and hysteresis is observed in the area surrounded by the gray line.
    } \label{fig:RandomNet_Pump}
\end{figure}

\section{Results}
\subsection{Randomly generated networks}
To understand the behaviors of the above model, we first randomly generated hundreds of intracellular reaction networks and numerically investigated the networks~\footnote{The reaction networks are randomly generated such that nutrient chemical $X_0$ can only be a substrate of intracellular reactions, transporter chemical $X_1$ cannot catalyze any intracellular reactions, and autocatalytic reactions do not occur}. 
We then observed discontinuous transitions between growth and dormant phases against external nutrient abundance $\Sext$. 

As an example, we consider the reaction network in Fig.~\ref{fig:RandomNet_Pump}(a). 
In Fig.~\ref{fig:RandomNet_Pump}(b), the steady growth rate $\mu^\ast$, numerically obtained by solving the dynamics~(\ref{eq:dxdt}-\ref{eq:dydt}), is plotted against the environmental nutrient concentration $\Sext$. 
As shown, $\mu^\ast$ decreases by orders of magnitude at $\Sext={}^\exists\Sext^c$, thus demonstrating the transition from growth to growth-arrested dormant phase. 
In addition, when $\Sext$ is increased starting from the dormant phase, the transition occurs at a larger $\Sext$, thus demonstrating hysteresis and bistability between the growth and dormant phases with intermediate levels of nutrient supply $\Sext$ (Fig.~\ref{fig:RandomNet_Pump}(b))~\footnote{
In this Letter, we basically plot the steady states that cells reach when $\Sext$ is lowered, i.e., the initial state for each set of parameters is given as the steady state with slightly higher $\Sext$. In actual cells, the fluctuation of internal concentrations leads to attractor selection~\cite{kashiwagi2006adaptive}; thus, the branch with the higher growth rate would be more likely to be observed.}, as is observed for real microbes~\cite{kotte2014phenotypic,veening2008bistability,krishna2018minimal}. 

Through this \textit{growth-dormant transition}, the intracellular chemical compositions and dominant reactions at work also change drastically (Fig.~\ref{fig:RandomNet_Pump}(b,c)). 
In the growth phase with larger $\Sext$, the nutrient influx is concentrated on an \textit{autocatalytic growth subnetwork}~\cite{eigen2012hypercycle,kauffman1993origins,jain1998autocatalytic,kaneko2005recursive,blokhuis2020universal} consisting of a few chemicals and reactions that connect the nutrient to the transporter (and associated byproducts). 
In contrast, in the dormant phase, fluxes spread over many chemicals in a subnetwork that cannot sustain growth by itself, which we term the \textit{non-growing subnetwork}. Here, a subnetwork is referred to as a closed set of chemicals and reactions whose catalysts, substrates, and products are included therein and in which no other chemicals are included (see also SM, Sec.~\ref{sec:random_net} for details). 
In contrast, in the dormant phase, fluxes spread over many chemicals in a subnetwork that cannot sustain growth by itself, which we term the \textit{non-growing subnetwork}. Here, a subnetwork is referred to as a closed set of chemicals and reactions whose catalysts and substrates are included therein (see also SM, Sec.~\ref{sec:random_net} for details). 
These subnetworks compete with each other while also overlapping: the activation of the autocatalytic growth subnetwork suppresses the non-growing subnetwork via growth-induced dilution, whereas the latter inhibits the former by the accumulation of complexes because reactants that are combined into some complexes in the non-growing subnetwork cannot be used for the reactions in the autocatalytic growth subnetwork. 
Consistently, the total concentration of complexes $Y := \sum_\rho y_\rho$ increases across the growth-dormant transition, as shown in Fig.~\ref{fig:RandomNet_Pump}(d). 
Due to the competition between the autocatalytic and non-growing subnetworks, this transition exhibits discontinuity, hysteresis, and bistability (Fig.~\ref{fig:RandomNet_Pump}(b)).

To measure such competition between autocatalytic and non-growing subnetworks, we defined \textit{composition entropy}, $H(\x,\y):=-\sum_ix_i\log x_i-\sum_\rho2y_\rho\log(2y_\rho)$. 
In general, in the growth phase, an autocatalytic subnetwork with the largest growth rate should be dominant and $H$ should be small, whereas in the dormant phase, many reactions and chemicals in the non-growing subnetwork could be engaged to the same degree and $H$ can be relatively large. 
As both subnetworks are comparably active near the critical nutrient concentration $\Sext^c$, the composition entropy $H$, or the diversity of the intracellular chemical composition, reaches a maximum near the transition point (Fig.~\ref{fig:RandomNet_Pump}(d)). 
Notably, such a trend is common among randomly generated networks (Fig.~\ref{fig:RandomNet_Pump_S1b}). From a biological perspective, this prediction would be consistent with the observations that stringent responses increase the diversity of the cellular components during the transition and in the dormant phase~\cite{navarro2010stationary,jaishankar2017molecular,nirupama2018stress}. 

The suppression of growth at the transition can be understood as a type of jamming caused by the accumulation of intermediate complexes: when the total concentration of complexes is larger, the free catalysts necessary for reactions in the autocatalytic growth subnetwork are limited~\cite{hatakeyama2017metabolic}. Consistently, with $v\,{>}\,^\exists v^c$, discontinuous transition and hysteresis are not observed against changes in $\Sext$ (Fig.~\ref{fig:RandomNet_Pump}(e)). 
Moreover, the dependence of the steady growth rate $\mu^\ast$ on $(\Sext,v)$ in Fig.~\ref{fig:RandomNet_Pump}(e) suggests a cusp bifurcation in the dynamical systems theory (as is also confirmed by the following mean-field analysis). 
We also found that as $v$ is smaller, both $\Sext^c$ and $\mu_{\max}$ are smaller; in other words, when $v$ varies, a trade-off occurs between maximum growth rate $\mu_{\max}$ and minimal nutrient concentration for the growth phase, $\Sext^c$. 
Such a trade-off has historically been considered a result of evolution leading to adaptations to either abundant or scarce nutrient environments~\cite{shipley1988relationship,fink2023microbial}, whereas our results suggest that this trade-off is a universal feature of growing cells with complex reaction networks.

Statistically, sufficiently large reaction networks are expected to include non-growing subnetwork(s) in addition to autocatalytic growth subnetwork(s). Indeed, even with $n=10\sim30$, about half of the randomly generated networks exhibited growth-dormant transitions (Fig.~\ref{fig:RandomNet_Pump_Stat}(a) in SM). 
In addition, the proportion of networks that exhibit transitions is maximal for relatively sparse reaction networks, and the peak value gradually increases as the number $n$ of chemicals increases. 
The following characteristics are also common to such networks: (i) growth-dormant transition against changes in $\Sext$ requires small $v$; 
(ii) hysteresis against changes in $\Sext$; 
(iii) increases in composition entropy $H$ around the transitions; and (iv) a trade-off between maximum growth rates $\mu_{\max}$ and minimal nutrient concentrations $\Sext^c$ to sustain growth. 
These results suggest the universality of growth-dormant transitions due to reactant competition via complex formation in complicated reaction networks, as is the case for metabolic networks in actual cells. 

\begin{figure}[tb]
    \centering 
    \includegraphics[width=\linewidth, clip]{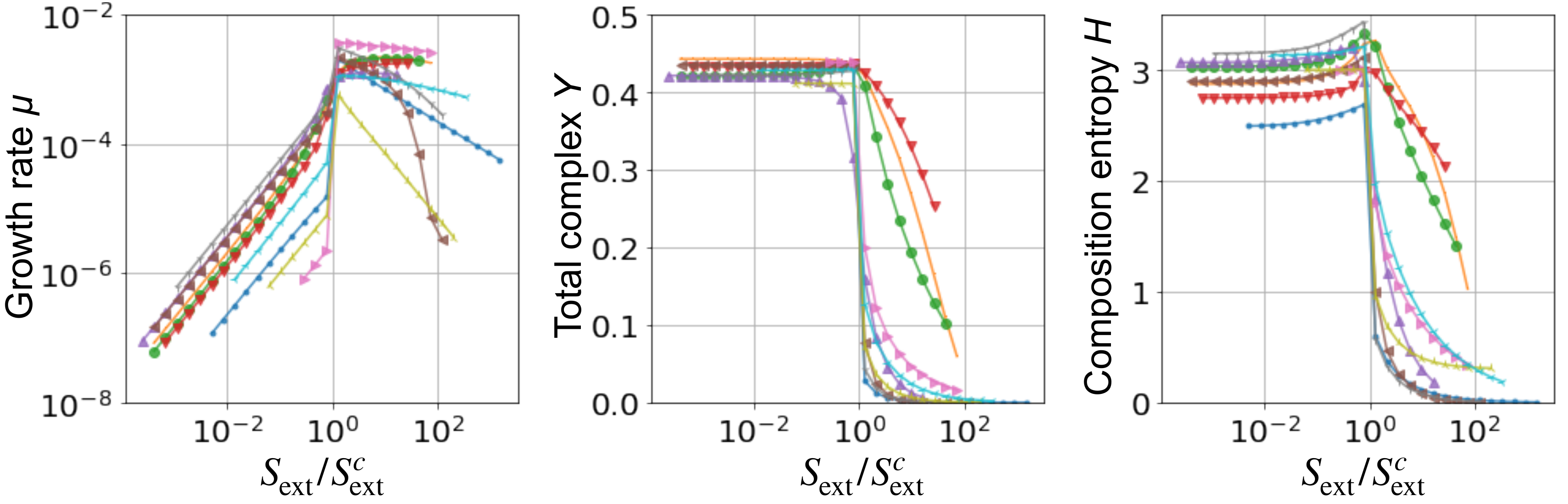}
    \caption{
    The steady growth rate $\mu^\ast$, the total concentration of complexes $Y$, and composition entropy $H$ are plotted against $\Sext/\Sext^c$ for several randomly generated networks that exhibit growth-dormant transitions in different colors. 
    $n=10,N_r=30$, $v=0.01$. 
    } \label{fig:RandomNet_Pump_S1b}
\end{figure}

We also numerically calculated the time for growth recovery after starvation as follows: 
First, up to $t=0$, cells are set in nutrient-rich conditions with sufficiently large $\Sext$, and they remain in steady states with exponential growth.
Second, the external nutrient supply is instantaneously depleted to $\Sext=0$ until $t=\Tst$. 
Finally, $\Sext$ is instantaneously increased to the original value. Then, a certain period $\Tlag\gg1/\mu_{\max}$, known as the lag time, is required for the cell to recover the original exponential growth, 
if the non-growing subnetwork is not a cycle and the amount of transporter chemical is sufficiently reduced therein; the lag time $\Tlag$ increases with starvation time $\Tst$ in the form $\Tlag\propto\Tst^\beta$ for a certain range (up to some saturation time), as observed in real microbes~\cite{augustin2000model,pin2008single} (see SM, Fig.~\ref{fig:RandomNet_DormantII} for an example). 
Here, the transporter's concentration gradually decreases under starvation; thus, the time for growth recovery, which requires the regain of the transporter, increases with starvation time~\footnote{If cells reach fixed points that completely lose the transporter(s), cells cannot recover growth. We term such steady states \textit{death phases} (see SM, Sec.~\ref{sec:random_net} for details).}. The exponent $\beta$ ranges from approximately $0.1$ to $0.5$ depending on the network structures that alter the intracellular reaction dynamics.

\begin{figure*}[tbh]
    \centering
    \includegraphics[width=\linewidth]{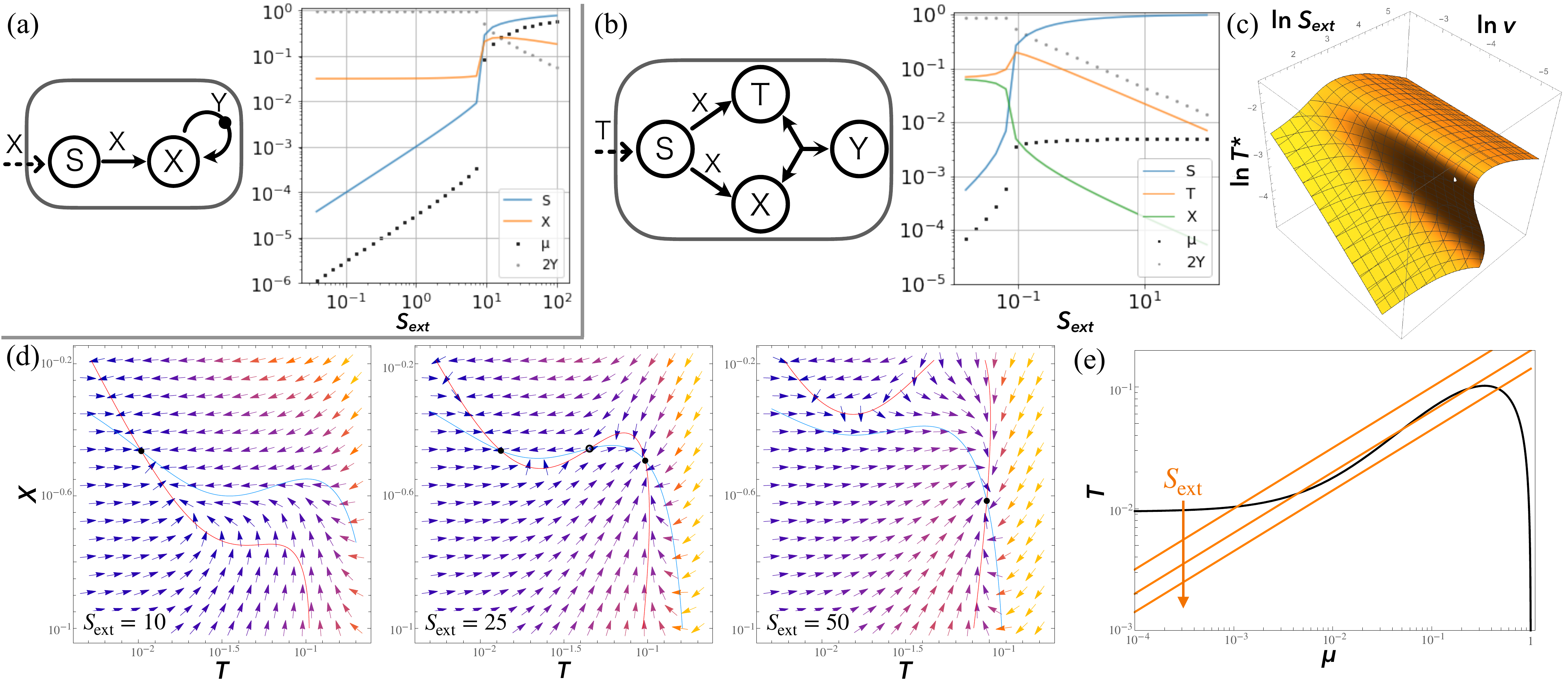}
    \caption{Mean-field models. 
    (a) Mean-field model with $S$, $X$, and $Y$ only. (left) Network structure. (right) Dependence of $\mu^\ast$ and steady states on $\Sext$. 
    $\alpha=3,v=0.001$. 
    (b-e) Mean-field model with the distinction between transporter $T$ and the remaining chemicals $X$. Unless otherwise stated, $v=0.01,n_X=2$. 
    (b) (left) Network structure. (right) Dependence of $\mu^\ast$ and steady states on $\Sext$. 
    (c) Bifurcation diagram: Dependence of $T^\ast$ on $(\Sext,v)$. 
    (d) Flow diagram in the phase space $(T,X)$. Red and blue lines represent $T$-nullcline and $X$-nullcline, respectively. 
    Arrows with brighter colors correspond to faster flows. 
    (e) Self-consistent equation for $T$ and $\mu$. 
    Black and orange lines depict $T=T^\ast(\mu;v,n_X)$ (Eq.~\eqref{eq:mean-field_simplest_T} in SM) and $T=\left(\mu/\Sext\right)^{1/\alpha}$ with $\Sext=10,25,50$, respectively. 
    } \label{fig:mean-field_main}
\end{figure*}

\subsection{Mean-field analysis} 
To further investigate the mechanism underlying the growth-dormant transition in terms of dynamical systems theory, we constructed mean-field models. 
First, we considered a model with one effective concentration variable $X$ and associated complexes $Y$ in addition to the nutrient $S$ (Fig.~\ref{fig:mean-field_main}(a)). This model exhibits the growth-dormant transition, although it requires extremely small values of the parameter $v$, $v<\mu_{\max}$, and $\alpha>2$.

Then, we considered another mean-field model that incorporates another variable $T$ representing the mean-field for the concentration of the transporter(s) in addition to $X$ representing the remaining non-nutrient chemicals (Fig.~\ref{fig:mean-field_main}(b)). The number of chemicals represented by $X$ and $T$ are denoted by $n_X$ and $n_T$, respectively. 
As only the complex $Y$ between $X$ and $T$ is considered for simplicity in this model~\footnote{Note that there are several possibilities for the choice of mean-field models with effective reactions among $X$ and $T$ that exhibit the growth-dormant transition (see SM, Sec.~\ref{sec:mean-field} for details).}, it includes the autocatalytic growth subnetwork, $S+X\to T+X$ and $S+X\to 2X$, and the single non-growing subnetwork, $T+X\rightleftharpoons Y$. 
This mean-field model reproduces common behaviors observed for randomly generated networks, including discontinuous growth-dormant transitions with $v > \mu_{\max}$ (Fig.~\ref{fig:mean-field_main}(b)). 
The transition occurs when $n_X > n_T=1$, and a larger number $n_X$ of $X$ leads to a larger $\Sext^c$ (see SM, Fig.~\ref{fig:mean-field_simplest}). 

From the bifurcation analysis (Fig.~\ref{fig:mean-field_main}(c-d)), we found that the growth-dormant transition occurs as a cusp bifurcation against changes in $\Sext$ and $v$~\cite{arnold2012geometrical}. This observation can explain some of the above-mentioned properties of randomly generated networks, i.e., discontinuous transitions and hysteresis. 
Notably, although both transporter $T$ and the remaining chemicals $X$ are essential for cell growth, their competition leads to a flow field with mutual inhibition as in the toggle switch at the intermediate value of $\Sext$. 
Furthermore, from the self-consistent equation for the steady growth rate $\mu^\ast$, we can determine where and how the growth-dormant transition occurs (Fig.~\ref{fig:mean-field_main}(e)).

\section{Discussion}
In this Letter, we studied a model of catalytic reaction networks wherein a variety of components react via the formation of intermediate complexes. This model exhibits discontinuous growth-dormant transitions against nutrient conditions as long as the formed complexes have sufficient lifetimes (i.e., with small $v$). 
This transition to growth-arrested dormant phases is caused by the accumulation of intermediate complexes under nutrient-poor conditions, which results in the jamming of reactions in the autocatalytic growth subnetwork and the relatively even distribution of chemical concentrations over diverse components in the non-growing subnetwork. 
Remarkably, other basic characteristics of dormancy, i.e., hysteresis between the exponential growth and dormant phases, the lag time for growth recovery after starvation, and a trade-off between maximum growth rate $\mu_{\max}$ and minimal nutrient concentration $\Sext^c$ to sustain growth (or a sort of sensitivity to nutrient scarcity) are also reproduced. 
These results indicate that growth-dormant transitions and dormancy might be inevitable for cells that grow via complex-forming catalytic reaction networks and likely emerge without tuning by evolution or adaptation; thus, even protocells at the primitive stage of life~\cite{luisi2016emergence,ameta2021self} are expected to exhibit such transitions to dormancy, which would be relevant to their survival under environmental stresses. 
On the other hand, further studies of detailed realistic models, such as those including distributed parameters and more realistic network structures, will be necessary to reveal how the above fundamental characteristics of dormancy are preserved or changed by evolution.

Moreover, the composition entropy $H$ is predicted to increase towards the transition point as a result of the competition between the autocatalytic and non-growing subnetworks. Biologically, it would correspond to the stringent responses that increase the diversity of the intracellular components~\cite{navarro2010stationary,jaishankar2017molecular,nirupama2018stress}. It will be crucial to experimentally verify the predicted increases in the composition entropy and jamming of reactions due to the accumulation of complexes. 

We also analyzed the dynamics of mean-field models and thereby demonstrate that the growth-dormant transition occurs as a cusp bifurcation~\cite{arnold2012geometrical}, which supports the discontinuity of the transitions as well as hysteresis. The validity of the coarse-grained mean-field models suggests the universality of the growth-dormant transition across many-body reaction systems. 
Note that, although the mean-field models in Fig.~\ref{fig:mean-field_main} capture the mechanism of the dormancy or growth arrest due to the accumulation of complexes, the diversification of components in nutrient-poor conditions owing to competition between subnetworks~\cite{kamimura2016negative} is not addressed. 
To consider such diversification and improve the mean-field theory, the incorporation of two-body correlations among the concentrations beyond their mean will be required. 

In conclusion, our study explains the ubiquity and fundamental characteristics of dormancy as general properties in reaction networks with complex formation, by offering a coherent view of cell growth and dormancy.

\begin{acknowledgments}
    The authors would like to thank Yuichi Wakamoto, Yusuke Himeoka, Tetsuhiro S. Hatakeyama, and Shuji Ishihara for stimulating discussions. J.~F.~Y.\ is supported by Grant-in-Aid for JSPS Fellows (21J22920). K.~K. is supported by Grant-in-Aid for Scientific Research (A) (20H00123) from the Ministry of Education, Culture, Sports, Science, and Technology (MEXT) of Japan and the Novo Nordisk Foundation. 
\end{acknowledgments}

\bibliographystyle{apsrev4-2}
\bibliography{bibliography}

\begin{thebibliography}{44}%
\makeatletter
\providecommand \@ifxundefined [1]{%
 \@ifx{#1\undefined}
}%
\providecommand \@ifnum [1]{%
 \ifnum #1\expandafter \@firstoftwo
 \else \expandafter \@secondoftwo
 \fi
}%
\providecommand \@ifx [1]{%
 \ifx #1\expandafter \@firstoftwo
 \else \expandafter \@secondoftwo
 \fi
}%
\providecommand \natexlab [1]{#1}%
\providecommand \enquote  [1]{``#1''}%
\providecommand \bibnamefont  [1]{#1}%
\providecommand \bibfnamefont [1]{#1}%
\providecommand \citenamefont [1]{#1}%
\providecommand \href@noop [0]{\@secondoftwo}%
\providecommand \href [0]{\begingroup \@sanitize@url \@href}%
\providecommand \@href[1]{\@@startlink{#1}\@@href}%
\providecommand \@@href[1]{\endgroup#1\@@endlink}%
\providecommand \@sanitize@url [0]{\catcode `\\12\catcode `\$12\catcode
  `\&12\catcode `\#12\catcode `\^12\catcode `\_12\catcode `\%12\relax}%
\providecommand \@@startlink[1]{}%
\providecommand \@@endlink[0]{}%
\providecommand \url  [0]{\begingroup\@sanitize@url \@url }%
\providecommand \@url [1]{\endgroup\@href {#1}{\urlprefix }}%
\providecommand \urlprefix  [0]{URL }%
\providecommand \Eprint [0]{\href }%
\providecommand \doibase [0]{https://doi.org/}%
\providecommand \selectlanguage [0]{\@gobble}%
\providecommand \bibinfo  [0]{\@secondoftwo}%
\providecommand \bibfield  [0]{\@secondoftwo}%
\providecommand \translation [1]{[#1]}%
\providecommand \BibitemOpen [0]{}%
\providecommand \bibitemStop [0]{}%
\providecommand \bibitemNoStop [0]{.\EOS\space}%
\providecommand \EOS [0]{\spacefactor3000\relax}%
\providecommand \BibitemShut  [1]{\csname bibitem#1\endcsname}%
\let\auto@bib@innerbib\@empty
\bibitem [{Note1()}]{Note1}%
  \BibitemOpen
  \bibinfo {note} {Such growth-arrested states are often termed dormant,
  non-growing, or quiescent phases/states, as well as stationary phase for an
  ensemble of cells. Herein, we adopt the term dormant phase}\BibitemShut
  {NoStop}%
\bibitem [{\citenamefont {Gray}\ \emph {et~al.}(2004)\citenamefont {Gray},
  \citenamefont {Petsko}, \citenamefont {Johnston}, \citenamefont {Ringe},
  \citenamefont {Singer},\ and\ \citenamefont
  {Werner-Washburne}}]{gray2004sleeping}%
  \BibitemOpen
  \bibfield  {author} {\bibinfo {author} {\bibfnamefont {J.~V.}\ \bibnamefont
  {Gray}}, \bibinfo {author} {\bibfnamefont {G.~A.}\ \bibnamefont {Petsko}},
  \bibinfo {author} {\bibfnamefont {G.~C.}\ \bibnamefont {Johnston}}, \bibinfo
  {author} {\bibfnamefont {D.}~\bibnamefont {Ringe}}, \bibinfo {author}
  {\bibfnamefont {R.~A.}\ \bibnamefont {Singer}},\ and\ \bibinfo {author}
  {\bibfnamefont {M.}~\bibnamefont {Werner-Washburne}},\ }\href@noop {}
  {\bibfield  {journal} {\bibinfo  {journal} {Microbiology and molecular
  biology reviews}\ }\textbf {\bibinfo {volume} {68}},\ \bibinfo {pages} {187}
  (\bibinfo {year} {2004})}\BibitemShut {NoStop}%
\bibitem [{\citenamefont {Finkel}(2006)}]{finkel2006long}%
  \BibitemOpen
  \bibfield  {author} {\bibinfo {author} {\bibfnamefont {S.~E.}\ \bibnamefont
  {Finkel}},\ }\href@noop {} {\bibfield  {journal} {\bibinfo  {journal} {Nature
  Reviews Microbiology}\ }\textbf {\bibinfo {volume} {4}},\ \bibinfo {pages}
  {113} (\bibinfo {year} {2006})}\BibitemShut {NoStop}%
\bibitem [{\citenamefont {Navarro~Llorens}\ \emph {et~al.}(2010)\citenamefont
  {Navarro~Llorens}, \citenamefont {Tormo},\ and\ \citenamefont
  {Mart{\'\i}nez-Garc{\'\i}a}}]{navarro2010stationary}%
  \BibitemOpen
  \bibfield  {author} {\bibinfo {author} {\bibfnamefont {J.~M.}\ \bibnamefont
  {Navarro~Llorens}}, \bibinfo {author} {\bibfnamefont {A.}~\bibnamefont
  {Tormo}},\ and\ \bibinfo {author} {\bibfnamefont {E.}~\bibnamefont
  {Mart{\'\i}nez-Garc{\'\i}a}},\ }\href@noop {} {\bibfield  {journal} {\bibinfo
   {journal} {FEMS microbiology reviews}\ }\textbf {\bibinfo {volume} {34}},\
  \bibinfo {pages} {476} (\bibinfo {year} {2010})}\BibitemShut {NoStop}%
\bibitem [{\citenamefont {Del~Giorgio}\ and\ \citenamefont
  {Gasol}(2008)}]{del2008physiological}%
  \BibitemOpen
  \bibfield  {author} {\bibinfo {author} {\bibfnamefont {P.~A.}\ \bibnamefont
  {Del~Giorgio}}\ and\ \bibinfo {author} {\bibfnamefont {J.~M.}\ \bibnamefont
  {Gasol}},\ }\href@noop {} {\bibfield  {journal} {\bibinfo  {journal}
  {Microbial ecology of the oceans}\ }\textbf {\bibinfo {volume} {2}},\
  \bibinfo {pages} {243} (\bibinfo {year} {2008})}\BibitemShut {NoStop}%
\bibitem [{\citenamefont {Gefen}\ \emph {et~al.}(2014)\citenamefont {Gefen},
  \citenamefont {Fridman}, \citenamefont {Ronin},\ and\ \citenamefont
  {Balaban}}]{gefen2014direct}%
  \BibitemOpen
  \bibfield  {author} {\bibinfo {author} {\bibfnamefont {O.}~\bibnamefont
  {Gefen}}, \bibinfo {author} {\bibfnamefont {O.}~\bibnamefont {Fridman}},
  \bibinfo {author} {\bibfnamefont {I.}~\bibnamefont {Ronin}},\ and\ \bibinfo
  {author} {\bibfnamefont {N.~Q.}\ \bibnamefont {Balaban}},\ }\href@noop {}
  {\bibfield  {journal} {\bibinfo  {journal} {Proceedings of the National
  Academy of Sciences}\ }\textbf {\bibinfo {volume} {111}},\ \bibinfo {pages}
  {556} (\bibinfo {year} {2014})}\BibitemShut {NoStop}%
\bibitem [{\citenamefont {Kirchman}(2018)}]{kirchman2018processes}%
  \BibitemOpen
  \bibfield  {author} {\bibinfo {author} {\bibfnamefont {D.~L.}\ \bibnamefont
  {Kirchman}},\ }\bibinfo {title} {Processes in microbial ecology}\ (\bibinfo
  {publisher} {Oxford University Press},\ \bibinfo {year} {2018})\
  Chap.~\bibinfo {chapter} {6}\BibitemShut {NoStop}%
\bibitem [{\citenamefont {Kotte}\ \emph {et~al.}(2014)\citenamefont {Kotte},
  \citenamefont {Volkmer}, \citenamefont {Radzikowski},\ and\ \citenamefont
  {Heinemann}}]{kotte2014phenotypic}%
  \BibitemOpen
  \bibfield  {author} {\bibinfo {author} {\bibfnamefont {O.}~\bibnamefont
  {Kotte}}, \bibinfo {author} {\bibfnamefont {B.}~\bibnamefont {Volkmer}},
  \bibinfo {author} {\bibfnamefont {J.~L.}\ \bibnamefont {Radzikowski}},\ and\
  \bibinfo {author} {\bibfnamefont {M.}~\bibnamefont {Heinemann}},\ }\href@noop
  {} {\bibfield  {journal} {\bibinfo  {journal} {Molecular systems biology}\
  }\textbf {\bibinfo {volume} {10}},\ \bibinfo {pages} {736} (\bibinfo {year}
  {2014})}\BibitemShut {NoStop}%
\bibitem [{\citenamefont {Veening}\ \emph {et~al.}(2008)\citenamefont
  {Veening}, \citenamefont {Smits},\ and\ \citenamefont
  {Kuipers}}]{veening2008bistability}%
  \BibitemOpen
  \bibfield  {author} {\bibinfo {author} {\bibfnamefont {J.-W.}\ \bibnamefont
  {Veening}}, \bibinfo {author} {\bibfnamefont {W.~K.}\ \bibnamefont {Smits}},\
  and\ \bibinfo {author} {\bibfnamefont {O.~P.}\ \bibnamefont {Kuipers}},\
  }\href@noop {} {\bibfield  {journal} {\bibinfo  {journal} {Annual review of
  microbiology}\ }\textbf {\bibinfo {volume} {62}},\ \bibinfo {pages} {193}
  (\bibinfo {year} {2008})}\BibitemShut {NoStop}%
\bibitem [{\citenamefont {Krishna}\ and\ \citenamefont
  {Laxman}(2018)}]{krishna2018minimal}%
  \BibitemOpen
  \bibfield  {author} {\bibinfo {author} {\bibfnamefont {S.}~\bibnamefont
  {Krishna}}\ and\ \bibinfo {author} {\bibfnamefont {S.}~\bibnamefont
  {Laxman}},\ }\href@noop {} {\bibfield  {journal} {\bibinfo  {journal}
  {Molecular biology of the cell}\ }\textbf {\bibinfo {volume} {29}},\ \bibinfo
  {pages} {2243} (\bibinfo {year} {2018})}\BibitemShut {NoStop}%
\bibitem [{\citenamefont {Levin-Reisman}\ \emph {et~al.}(2010)\citenamefont
  {Levin-Reisman}, \citenamefont {Gefen}, \citenamefont {Fridman},
  \citenamefont {Ronin}, \citenamefont {Shwa}, \citenamefont {Sheftel},\ and\
  \citenamefont {Balaban}}]{levin2010automated}%
  \BibitemOpen
  \bibfield  {author} {\bibinfo {author} {\bibfnamefont {I.}~\bibnamefont
  {Levin-Reisman}}, \bibinfo {author} {\bibfnamefont {O.}~\bibnamefont
  {Gefen}}, \bibinfo {author} {\bibfnamefont {O.}~\bibnamefont {Fridman}},
  \bibinfo {author} {\bibfnamefont {I.}~\bibnamefont {Ronin}}, \bibinfo
  {author} {\bibfnamefont {D.}~\bibnamefont {Shwa}}, \bibinfo {author}
  {\bibfnamefont {H.}~\bibnamefont {Sheftel}},\ and\ \bibinfo {author}
  {\bibfnamefont {N.~Q.}\ \bibnamefont {Balaban}},\ }\href@noop {} {\bibfield
  {journal} {\bibinfo  {journal} {Nature methods}\ }\textbf {\bibinfo {volume}
  {7}},\ \bibinfo {pages} {737} (\bibinfo {year} {2010})}\BibitemShut {NoStop}%
\bibitem [{\citenamefont {J{\~o}ers}\ and\ \citenamefont
  {Tenson}(2016)}]{joers2016growth}%
  \BibitemOpen
  \bibfield  {author} {\bibinfo {author} {\bibfnamefont {A.}~\bibnamefont
  {J{\~o}ers}}\ and\ \bibinfo {author} {\bibfnamefont {T.}~\bibnamefont
  {Tenson}},\ }\href@noop {} {\bibfield  {journal} {\bibinfo  {journal}
  {Scientific reports}\ }\textbf {\bibinfo {volume} {6}},\ \bibinfo {pages} {1}
  (\bibinfo {year} {2016})}\BibitemShut {NoStop}%
\bibitem [{\citenamefont {Kaplan}\ \emph {et~al.}(2021)\citenamefont {Kaplan},
  \citenamefont {Reich}, \citenamefont {Oster}, \citenamefont {Maoz},
  \citenamefont {Levin-Reisman}, \citenamefont {Ronin}, \citenamefont {Gefen},
  \citenamefont {Agam},\ and\ \citenamefont {Balaban}}]{kaplan2021observation}%
  \BibitemOpen
  \bibfield  {author} {\bibinfo {author} {\bibfnamefont {Y.}~\bibnamefont
  {Kaplan}}, \bibinfo {author} {\bibfnamefont {S.}~\bibnamefont {Reich}},
  \bibinfo {author} {\bibfnamefont {E.}~\bibnamefont {Oster}}, \bibinfo
  {author} {\bibfnamefont {S.}~\bibnamefont {Maoz}}, \bibinfo {author}
  {\bibfnamefont {I.}~\bibnamefont {Levin-Reisman}}, \bibinfo {author}
  {\bibfnamefont {I.}~\bibnamefont {Ronin}}, \bibinfo {author} {\bibfnamefont
  {O.}~\bibnamefont {Gefen}}, \bibinfo {author} {\bibfnamefont
  {O.}~\bibnamefont {Agam}},\ and\ \bibinfo {author} {\bibfnamefont {N.~Q.}\
  \bibnamefont {Balaban}},\ }\href@noop {} {\bibfield  {journal} {\bibinfo
  {journal} {Nature}\ }\textbf {\bibinfo {volume} {600}},\ \bibinfo {pages}
  {290} (\bibinfo {year} {2021})}\BibitemShut {NoStop}%
\bibitem [{\citenamefont {Scott}\ and\ \citenamefont
  {Hwa}(2011)}]{scott2011bacterial}%
  \BibitemOpen
  \bibfield  {author} {\bibinfo {author} {\bibfnamefont {M.}~\bibnamefont
  {Scott}}\ and\ \bibinfo {author} {\bibfnamefont {T.}~\bibnamefont {Hwa}},\
  }\href@noop {} {\bibfield  {journal} {\bibinfo  {journal} {Current opinion in
  biotechnology}\ }\textbf {\bibinfo {volume} {22}},\ \bibinfo {pages} {559}
  (\bibinfo {year} {2011})}\BibitemShut {NoStop}%
\bibitem [{\citenamefont {Kaneko}\ \emph {et~al.}(2015)\citenamefont {Kaneko},
  \citenamefont {Furusawa},\ and\ \citenamefont {Yomo}}]{kaneko2015universal}%
  \BibitemOpen
  \bibfield  {author} {\bibinfo {author} {\bibfnamefont {K.}~\bibnamefont
  {Kaneko}}, \bibinfo {author} {\bibfnamefont {C.}~\bibnamefont {Furusawa}},\
  and\ \bibinfo {author} {\bibfnamefont {T.}~\bibnamefont {Yomo}},\ }\href@noop
  {} {\bibfield  {journal} {\bibinfo  {journal} {Physical Review X}\ }\textbf
  {\bibinfo {volume} {5}},\ \bibinfo {pages} {011014} (\bibinfo {year}
  {2015})}\BibitemShut {NoStop}%
\bibitem [{\citenamefont {Jun}\ \emph {et~al.}(2018)\citenamefont {Jun},
  \citenamefont {Si}, \citenamefont {Pugatch},\ and\ \citenamefont
  {Scott}}]{jun2018fundamental}%
  \BibitemOpen
  \bibfield  {author} {\bibinfo {author} {\bibfnamefont {S.}~\bibnamefont
  {Jun}}, \bibinfo {author} {\bibfnamefont {F.}~\bibnamefont {Si}}, \bibinfo
  {author} {\bibfnamefont {R.}~\bibnamefont {Pugatch}},\ and\ \bibinfo {author}
  {\bibfnamefont {M.}~\bibnamefont {Scott}},\ }\href@noop {} {\bibfield
  {journal} {\bibinfo  {journal} {Reports on Progress in Physics}\ }\textbf
  {\bibinfo {volume} {81}},\ \bibinfo {pages} {056601} (\bibinfo {year}
  {2018})}\BibitemShut {NoStop}%
\bibitem [{\citenamefont {Jaishankar}\ and\ \citenamefont
  {Srivastava}(2017)}]{jaishankar2017molecular}%
  \BibitemOpen
  \bibfield  {author} {\bibinfo {author} {\bibfnamefont {J.}~\bibnamefont
  {Jaishankar}}\ and\ \bibinfo {author} {\bibfnamefont {P.}~\bibnamefont
  {Srivastava}},\ }\href@noop {} {\bibfield  {journal} {\bibinfo  {journal}
  {Frontiers in microbiology}\ }\textbf {\bibinfo {volume} {8}},\ \bibinfo
  {pages} {2000} (\bibinfo {year} {2017})}\BibitemShut {NoStop}%
\bibitem [{\citenamefont {Himeoka}\ and\ \citenamefont
  {Kaneko}(2017)}]{himeoka2017theory}%
  \BibitemOpen
  \bibfield  {author} {\bibinfo {author} {\bibfnamefont {Y.}~\bibnamefont
  {Himeoka}}\ and\ \bibinfo {author} {\bibfnamefont {K.}~\bibnamefont
  {Kaneko}},\ }\href@noop {} {\bibfield  {journal} {\bibinfo  {journal}
  {Physical Review X}\ }\textbf {\bibinfo {volume} {7}},\ \bibinfo {pages}
  {021049} (\bibinfo {year} {2017})}\BibitemShut {NoStop}%
\bibitem [{\citenamefont {Reich}\ \emph {et~al.}(2022)\citenamefont {Reich},
  \citenamefont {Maoz}, \citenamefont {Kaplan}, \citenamefont {Rappeport},
  \citenamefont {Balaban},\ and\ \citenamefont {Agam}}]{reich2022slow}%
  \BibitemOpen
  \bibfield  {author} {\bibinfo {author} {\bibfnamefont {S.}~\bibnamefont
  {Reich}}, \bibinfo {author} {\bibfnamefont {S.}~\bibnamefont {Maoz}},
  \bibinfo {author} {\bibfnamefont {Y.}~\bibnamefont {Kaplan}}, \bibinfo
  {author} {\bibfnamefont {H.}~\bibnamefont {Rappeport}}, \bibinfo {author}
  {\bibfnamefont {N.}~\bibnamefont {Balaban}},\ and\ \bibinfo {author}
  {\bibfnamefont {O.}~\bibnamefont {Agam}},\ }\href@noop {} {\bibfield
  {journal} {\bibinfo  {journal} {Physical Review Research}\ }\textbf {\bibinfo
  {volume} {4}},\ \bibinfo {pages} {033127} (\bibinfo {year}
  {2022})}\BibitemShut {NoStop}%
\bibitem [{\citenamefont {Furusawa}\ and\ \citenamefont
  {Kaneko}(2003)}]{furusawa2003zipf}%
  \BibitemOpen
  \bibfield  {author} {\bibinfo {author} {\bibfnamefont {C.}~\bibnamefont
  {Furusawa}}\ and\ \bibinfo {author} {\bibfnamefont {K.}~\bibnamefont
  {Kaneko}},\ }\href@noop {} {\bibfield  {journal} {\bibinfo  {journal}
  {Physical Review Letters}\ }\textbf {\bibinfo {volume} {90}},\ \bibinfo
  {pages} {088102} (\bibinfo {year} {2003})}\BibitemShut {NoStop}%
\bibitem [{\citenamefont {Furusawa}\ and\ \citenamefont
  {Kaneko}(2012)}]{furusawa2012adaptation}%
  \BibitemOpen
  \bibfield  {author} {\bibinfo {author} {\bibfnamefont {C.}~\bibnamefont
  {Furusawa}}\ and\ \bibinfo {author} {\bibfnamefont {K.}~\bibnamefont
  {Kaneko}},\ }\href@noop {} {\bibfield  {journal} {\bibinfo  {journal}
  {Physical Review Letters}\ }\textbf {\bibinfo {volume} {108}},\ \bibinfo
  {pages} {208103} (\bibinfo {year} {2012})}\BibitemShut {NoStop}%
\bibitem [{Note2()}]{Note2}%
  \BibitemOpen
  \bibinfo {note} {The reaction networks are randomly generated such that
  nutrient chemical $X_0$ can only be a substrate of intracellular reactions,
  transporter chemical $X_1$ cannot catalyze any intracellular reactions, and
  autocatalytic reactions do not occur}\BibitemShut {NoStop}%
\bibitem [{Note3()}]{Note3}%
  \BibitemOpen
  \bibinfo {note} {In this Letter, we basically plot the steady states that
  cells reach when $S_\protect \mathrm {ext}$ is lowered, i.e., the initial
  state for each set of parameters is given as the steady state with slightly
  higher $S_\protect \mathrm {ext}$. In actual cells, the fluctuation of
  internal concentrations leads to attractor selection~\cite
  {kashiwagi2006adaptive}; thus, the branch with the higher growth rate would
  be more likely to be observed.}\BibitemShut {Stop}%
\bibitem [{\citenamefont {Eigen}\ and\ \citenamefont
  {Schuster}(2012)}]{eigen2012hypercycle}%
  \BibitemOpen
  \bibfield  {author} {\bibinfo {author} {\bibfnamefont {M.}~\bibnamefont
  {Eigen}}\ and\ \bibinfo {author} {\bibfnamefont {P.}~\bibnamefont
  {Schuster}},\ }\href@noop {} {\emph {\bibinfo {title} {The hypercycle: a
  principle of natural self-organization}}}\ (\bibinfo  {publisher} {Springer
  Science \& Business Media},\ \bibinfo {year} {2012})\BibitemShut {NoStop}%
\bibitem [{\citenamefont {Kauffman}\ \emph {et~al.}(1993)\citenamefont
  {Kauffman} \emph {et~al.}}]{kauffman1993origins}%
  \BibitemOpen
  \bibfield  {author} {\bibinfo {author} {\bibfnamefont {S.~A.}\ \bibnamefont
  {Kauffman}} \emph {et~al.},\ }\href@noop {} {\emph {\bibinfo {title} {The
  origins of order: Self-organization and selection in evolution}}}\ (\bibinfo
  {publisher} {Oxford University Press, USA},\ \bibinfo {year}
  {1993})\BibitemShut {NoStop}%
\bibitem [{\citenamefont {Jain}\ and\ \citenamefont
  {Krishna}(1998)}]{jain1998autocatalytic}%
  \BibitemOpen
  \bibfield  {author} {\bibinfo {author} {\bibfnamefont {S.}~\bibnamefont
  {Jain}}\ and\ \bibinfo {author} {\bibfnamefont {S.}~\bibnamefont {Krishna}},\
  }\href@noop {} {\bibfield  {journal} {\bibinfo  {journal} {Physical Review
  Letters}\ }\textbf {\bibinfo {volume} {81}},\ \bibinfo {pages} {5684}
  (\bibinfo {year} {1998})}\BibitemShut {NoStop}%
\bibitem [{\citenamefont {Kaneko}(2005)}]{kaneko2005recursive}%
  \BibitemOpen
  \bibfield  {author} {\bibinfo {author} {\bibfnamefont {K.}~\bibnamefont
  {Kaneko}},\ }\href@noop {} {\bibfield  {journal} {\bibinfo  {journal}
  {Advances in Chemical Physics}\ }\textbf {\bibinfo {volume} {130}},\ \bibinfo
  {pages} {543} (\bibinfo {year} {2005})}\BibitemShut {NoStop}%
\bibitem [{\citenamefont {Blokhuis}\ \emph {et~al.}(2020)\citenamefont
  {Blokhuis}, \citenamefont {Lacoste},\ and\ \citenamefont
  {Nghe}}]{blokhuis2020universal}%
  \BibitemOpen
  \bibfield  {author} {\bibinfo {author} {\bibfnamefont {A.}~\bibnamefont
  {Blokhuis}}, \bibinfo {author} {\bibfnamefont {D.}~\bibnamefont {Lacoste}},\
  and\ \bibinfo {author} {\bibfnamefont {P.}~\bibnamefont {Nghe}},\ }\href@noop
  {} {\bibfield  {journal} {\bibinfo  {journal} {Proceedings of the National
  Academy of Sciences}\ }\textbf {\bibinfo {volume} {117}},\ \bibinfo {pages}
  {25230} (\bibinfo {year} {2020})}\BibitemShut {NoStop}%
\bibitem [{\citenamefont {Nirupama}\ \emph {et~al.}(2018)\citenamefont
  {Nirupama}, \citenamefont {Rajaraman},\ and\ \citenamefont
  {Yajurvedi}}]{nirupama2018stress}%
  \BibitemOpen
  \bibfield  {author} {\bibinfo {author} {\bibfnamefont {R.}~\bibnamefont
  {Nirupama}}, \bibinfo {author} {\bibfnamefont {B.}~\bibnamefont
  {Rajaraman}},\ and\ \bibinfo {author} {\bibfnamefont {H.}~\bibnamefont
  {Yajurvedi}},\ }\href@noop {} {\bibfield  {journal} {\bibinfo  {journal}
  {Imaging J Clin Med Sci}\ }\textbf {\bibinfo {volume} {5}},\ \bibinfo {pages}
  {8} (\bibinfo {year} {2018})}\BibitemShut {NoStop}%
\bibitem [{\citenamefont {Hatakeyama}\ and\ \citenamefont
  {Furusawa}(2017)}]{hatakeyama2017metabolic}%
  \BibitemOpen
  \bibfield  {author} {\bibinfo {author} {\bibfnamefont {T.~S.}\ \bibnamefont
  {Hatakeyama}}\ and\ \bibinfo {author} {\bibfnamefont {C.}~\bibnamefont
  {Furusawa}},\ }\href@noop {} {\bibfield  {journal} {\bibinfo  {journal} {PLoS
  computational biology}\ }\textbf {\bibinfo {volume} {13}},\ \bibinfo {pages}
  {e1005847} (\bibinfo {year} {2017})}\BibitemShut {NoStop}%
\bibitem [{\citenamefont {Shipley}\ and\ \citenamefont
  {Keddy}(1988)}]{shipley1988relationship}%
  \BibitemOpen
  \bibfield  {author} {\bibinfo {author} {\bibfnamefont {B.}~\bibnamefont
  {Shipley}}\ and\ \bibinfo {author} {\bibfnamefont {P.}~\bibnamefont
  {Keddy}},\ }\href@noop {} {\bibfield  {journal} {\bibinfo  {journal} {The
  Journal of Ecology}\ ,\ \bibinfo {pages} {1101}} (\bibinfo {year}
  {1988})}\BibitemShut {NoStop}%
\bibitem [{\citenamefont {Fink}\ \emph {et~al.}(2023)\citenamefont {Fink},
  \citenamefont {Held},\ and\ \citenamefont {Manhart}}]{fink2023microbial}%
  \BibitemOpen
  \bibfield  {author} {\bibinfo {author} {\bibfnamefont {J.~W.}\ \bibnamefont
  {Fink}}, \bibinfo {author} {\bibfnamefont {N.~A.}\ \bibnamefont {Held}},\
  and\ \bibinfo {author} {\bibfnamefont {M.}~\bibnamefont {Manhart}},\
  }\href@noop {} {\bibfield  {journal} {\bibinfo  {journal} {Proceedings of the
  National Academy of Sciences}\ }\textbf {\bibinfo {volume} {120}},\ \bibinfo
  {pages} {e2207295120} (\bibinfo {year} {2023})}\BibitemShut {NoStop}%
\bibitem [{\citenamefont {Augustin}\ \emph {et~al.}(2000)\citenamefont
  {Augustin}, \citenamefont {Rosso},\ and\ \citenamefont
  {Carlier}}]{augustin2000model}%
  \BibitemOpen
  \bibfield  {author} {\bibinfo {author} {\bibfnamefont {J.-C.}\ \bibnamefont
  {Augustin}}, \bibinfo {author} {\bibfnamefont {L.}~\bibnamefont {Rosso}},\
  and\ \bibinfo {author} {\bibfnamefont {V.}~\bibnamefont {Carlier}},\
  }\href@noop {} {\bibfield  {journal} {\bibinfo  {journal} {International
  Journal of food microbiology}\ }\textbf {\bibinfo {volume} {57}},\ \bibinfo
  {pages} {169} (\bibinfo {year} {2000})}\BibitemShut {NoStop}%
\bibitem [{\citenamefont {Pin}\ and\ \citenamefont
  {Baranyi}(2008)}]{pin2008single}%
  \BibitemOpen
  \bibfield  {author} {\bibinfo {author} {\bibfnamefont {C.}~\bibnamefont
  {Pin}}\ and\ \bibinfo {author} {\bibfnamefont {J.}~\bibnamefont {Baranyi}},\
  }\href@noop {} {\bibfield  {journal} {\bibinfo  {journal} {Applied and
  environmental microbiology}\ }\textbf {\bibinfo {volume} {74}},\ \bibinfo
  {pages} {2534} (\bibinfo {year} {2008})}\BibitemShut {NoStop}%
\bibitem [{Note4()}]{Note4}%
  \BibitemOpen
  \bibinfo {note} {If cells reach fixed points that completely lose the
  transporter(s), cells cannot recover growth. We term such steady states
  \protect \textit {death phases} (see SM, Sec.~\ref {sec:random_net} for
  details).}\BibitemShut {Stop}%
\bibitem [{Note5()}]{Note5}%
  \BibitemOpen
  \bibinfo {note} {Note that there are several possibilities for the choice of
  mean-field models with effective reactions among $X$ and $T$ that exhibit the
  growth-dormant transition (see SM, Sec.~\ref {sec:mean-field} for
  details).}\BibitemShut {Stop}%
\bibitem [{\citenamefont {Arnold}(2012)}]{arnold2012geometrical}%
  \BibitemOpen
  \bibfield  {author} {\bibinfo {author} {\bibfnamefont {V.~I.}\ \bibnamefont
  {Arnold}},\ }\href@noop {} {\emph {\bibinfo {title} {Geometrical methods in
  the theory of ordinary differential equations}}},\ Vol.\ \bibinfo {volume}
  {250}\ (\bibinfo  {publisher} {Springer Science \& Business Media},\ \bibinfo
  {year} {2012})\BibitemShut {NoStop}%
\bibitem [{\citenamefont {Luisi}(2016)}]{luisi2016emergence}%
  \BibitemOpen
  \bibfield  {author} {\bibinfo {author} {\bibfnamefont {P.~L.}\ \bibnamefont
  {Luisi}},\ }\href@noop {} {\emph {\bibinfo {title} {The emergence of life:
  from chemical origins to synthetic biology}}}\ (\bibinfo  {publisher}
  {Cambridge University Press},\ \bibinfo {year} {2016})\BibitemShut {NoStop}%
\bibitem [{\citenamefont {Ameta}\ \emph {et~al.}(2021)\citenamefont {Ameta},
  \citenamefont {Matsubara}, \citenamefont {Chakraborty}, \citenamefont
  {Krishna},\ and\ \citenamefont {Thutupalli}}]{ameta2021self}%
  \BibitemOpen
  \bibfield  {author} {\bibinfo {author} {\bibfnamefont {S.}~\bibnamefont
  {Ameta}}, \bibinfo {author} {\bibfnamefont {Y.~J.}\ \bibnamefont
  {Matsubara}}, \bibinfo {author} {\bibfnamefont {N.}~\bibnamefont
  {Chakraborty}}, \bibinfo {author} {\bibfnamefont {S.}~\bibnamefont
  {Krishna}},\ and\ \bibinfo {author} {\bibfnamefont {S.}~\bibnamefont
  {Thutupalli}},\ }\href@noop {} {\bibfield  {journal} {\bibinfo  {journal}
  {Life}\ }\textbf {\bibinfo {volume} {11}},\ \bibinfo {pages} {308} (\bibinfo
  {year} {2021})}\BibitemShut {NoStop}%
\bibitem [{\citenamefont {Kamimura}\ and\ \citenamefont
  {Kaneko}(2016)}]{kamimura2016negative}%
  \BibitemOpen
  \bibfield  {author} {\bibinfo {author} {\bibfnamefont {A.}~\bibnamefont
  {Kamimura}}\ and\ \bibinfo {author} {\bibfnamefont {K.}~\bibnamefont
  {Kaneko}},\ }\href@noop {} {\bibfield  {journal} {\bibinfo  {journal}
  {Physical Review E}\ }\textbf {\bibinfo {volume} {93}},\ \bibinfo {pages}
  {062419} (\bibinfo {year} {2016})}\BibitemShut {NoStop}%
\bibitem [{\citenamefont {Kashiwagi}\ \emph {et~al.}(2006)\citenamefont
  {Kashiwagi}, \citenamefont {Urabe}, \citenamefont {Kaneko},\ and\
  \citenamefont {Yomo}}]{kashiwagi2006adaptive}%
  \BibitemOpen
  \bibfield  {author} {\bibinfo {author} {\bibfnamefont {A.}~\bibnamefont
  {Kashiwagi}}, \bibinfo {author} {\bibfnamefont {I.}~\bibnamefont {Urabe}},
  \bibinfo {author} {\bibfnamefont {K.}~\bibnamefont {Kaneko}},\ and\ \bibinfo
  {author} {\bibfnamefont {T.}~\bibnamefont {Yomo}},\ }\href@noop {} {\bibfield
   {journal} {\bibinfo  {journal} {PloS One}\ }\textbf {\bibinfo {volume}
  {1}},\ \bibinfo {pages} {e49} (\bibinfo {year} {2006})}\BibitemShut {NoStop}%
\bibitem [{\citenamefont {de~Groot}\ \emph {et~al.}(2020)\citenamefont
  {de~Groot}, \citenamefont {Hulshof}, \citenamefont {Teusink}, \citenamefont
  {Bruggeman},\ and\ \citenamefont {Planqu{\'e}}}]{de2020elementary}%
  \BibitemOpen
  \bibfield  {author} {\bibinfo {author} {\bibfnamefont {D.~H.}\ \bibnamefont
  {de~Groot}}, \bibinfo {author} {\bibfnamefont {J.}~\bibnamefont {Hulshof}},
  \bibinfo {author} {\bibfnamefont {B.}~\bibnamefont {Teusink}}, \bibinfo
  {author} {\bibfnamefont {F.~J.}\ \bibnamefont {Bruggeman}},\ and\ \bibinfo
  {author} {\bibfnamefont {R.}~\bibnamefont {Planqu{\'e}}},\ }\href@noop {}
  {\bibfield  {journal} {\bibinfo  {journal} {PLoS computational biology}\
  }\textbf {\bibinfo {volume} {16}},\ \bibinfo {pages} {e1007559} (\bibinfo
  {year} {2020})}\BibitemShut {NoStop}%
\bibitem [{\citenamefont {M{\"u}ller}(2021)}]{muller2021elementary}%
  \BibitemOpen
  \bibfield  {author} {\bibinfo {author} {\bibfnamefont {S.}~\bibnamefont
  {M{\"u}ller}},\ }\href@noop {} {\bibfield  {journal} {\bibinfo  {journal}
  {bioRxiv}\ } (\bibinfo {year} {2021})}\BibitemShut {NoStop}%
\bibitem [{\citenamefont {Schilling}\ \emph {et~al.}(2000)\citenamefont
  {Schilling}, \citenamefont {Letscher},\ and\ \citenamefont
  {Palsson}}]{schilling2000theory}%
  \BibitemOpen
  \bibfield  {author} {\bibinfo {author} {\bibfnamefont {C.~H.}\ \bibnamefont
  {Schilling}}, \bibinfo {author} {\bibfnamefont {D.}~\bibnamefont
  {Letscher}},\ and\ \bibinfo {author} {\bibfnamefont {B.~{\O}.}\ \bibnamefont
  {Palsson}},\ }\href@noop {} {\bibfield  {journal} {\bibinfo  {journal}
  {Journal of theoretical biology}\ }\textbf {\bibinfo {volume} {203}},\
  \bibinfo {pages} {229} (\bibinfo {year} {2000})}\BibitemShut {NoStop}%
\end{thebibliography}%

\clearpage
\onecolumngrid
\appendix
\setcounter{figure}{0}
\renewcommand{\thesubsection}{\Alph{section}.\arabic{subsection}}
\renewcommand{\thefigure}{S\arabic{figure}}
\renewcommand{\thetable}{S\arabic{table}}
\setcounter{page}{1}

\begin{center}
    {\large\textbf{Supplemental Material} for
    \\
    \textbf{Universal Transitions between Growth and Dormancy via Intermediate Complex Formation}
    }
\end{center}

\begin{figure}[tbh]
    \centering 
    \includegraphics[width=0.95\linewidth, clip]{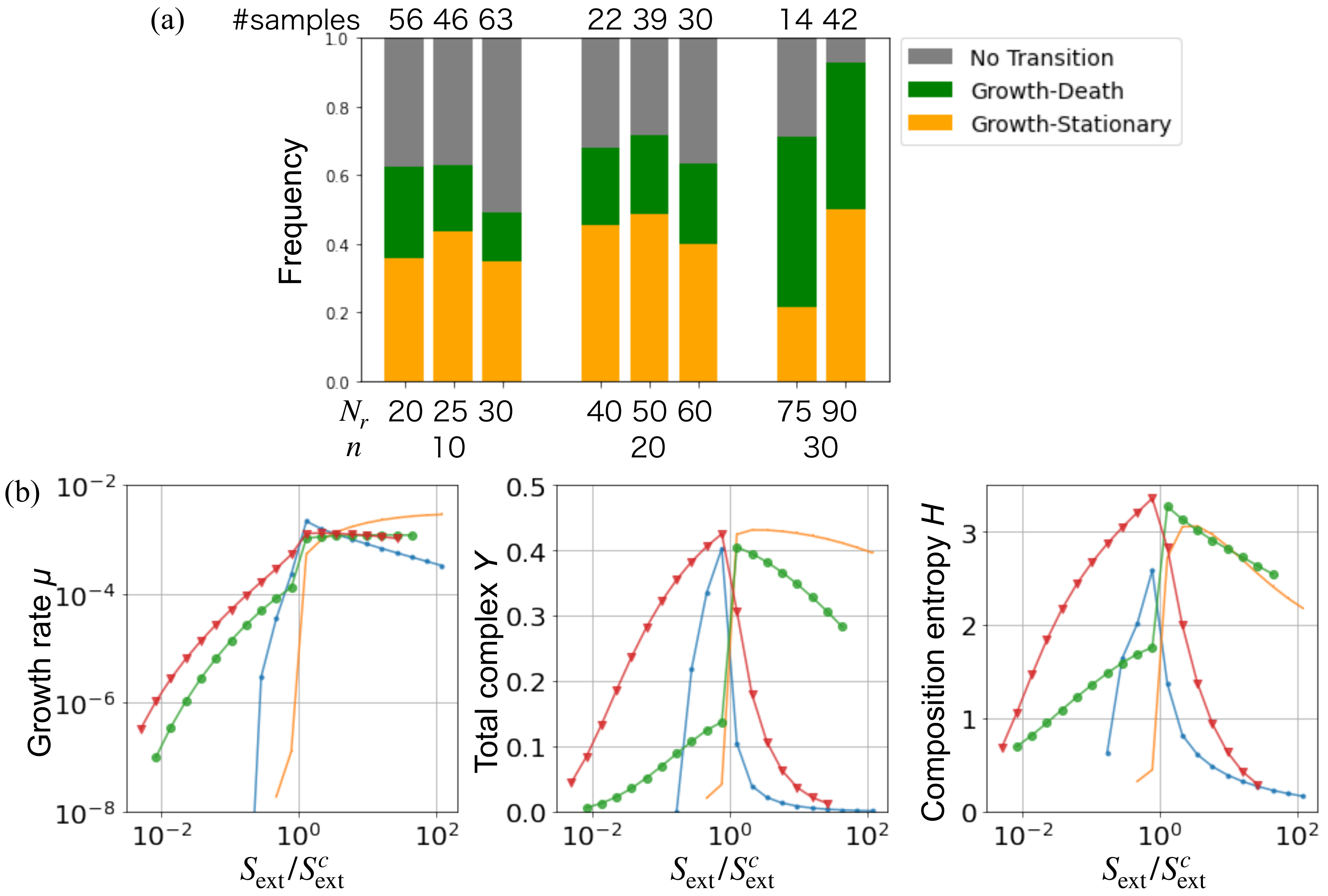}
    \caption{
    Statistics for randomly-generated networks. 
    (a) Fraction of randomly-generated networks exhibiting growth-dormant(-death) transition, growth-death transition, and no transition. 	
    Networks are randomly generated for each set of parameters (i.e., given $N_r$ and $n$ shown below). 
            (b) $\mu$, $Y$, and $H$ 
    are plotted against $\Sext/\Sext^c$ for networks showing growth-dormant transitions with non-growing subnetworks (II) that cannot be active exactly with $\Sext=0$. 
    $n=10,N_r=30$, $v=0.01$. 
    Different colors correspond to different networks. 
    See Fig.~\ref{fig:RandomNet_Pump_S1b} in the main text for networks with non-growing subnetworks (I) that are active even with $\Sext=0$. See also Sec.~\ref{sec:random_net} for details. 
    } \label{fig:RandomNet_Pump_Stat}
\end{figure}

\section{Details about complex-formation reaction-network model and numerical simulations}\label{sec:random_net}
\subsection{Details about phases and subnetworks}
Below we discuss the definition and characteristics of {the growth, dormant, and death phases and the subnetworks working in those phases.} 
Here, we referred to a subnetwork as a closed set of reactions and their catalysts, substrates, and products.

\textit{Growth phase.---}
Under the growth phase, a minimal autocatalytic subnetwork~\cite{eigen2012hypercycle,kauffman1993origins,jain1998autocatalytic,kaneko2005recursive,blokhuis2020universal} (and its byproducts) becomes dominant, which we term a autocatalytic growth subnetwork. 
It consumes nutrient chemical $X_0$ and produces transporter chemical $X_1$ (i.e., one of so-called elementary growth modes~\cite{de2020elementary,muller2021elementary}). 

\textit{Dormant phase.---}
Under the dormant phase, a closed set of chemicals and reactions that does not sustain growth by itself is dominantly active, which we term a non-growing subnetwork. 
The non-growing subnetworks are classified into two types, depending on whether they are cycles or not. 

Non-growing subnetworks (I) are {cycles within which the set of chemicals remains conserved in number (see Figs.~1~and~3 in the main text and Fig.~\ref{fig:mean-field_simplest} for other examples). Those subnetworks can be active even when $\Sext=0$. 
Such cycles are also termed \textit{allocatalytic cycles} (e.g., $A+E\leftrightarrows B+E$)~\cite{blokhuis2020universal}.} 

Non-growing subnetworks (II) are non-cyclic subnetworks ``parasitic'' to the autocatalytic growth subnetwork: For the reactions in those subnetworks to be sustained, some chemical(s) in the autocatalytic growth subnetwork are required as substrate(s) (see Fig.~\ref{fig:RandomNet_DormantII} for an example); thus, those subnetworks cannot be active when $\Sext=0$. 
They include some elementary flux modes~\cite{schilling2000theory} (i.e., elementary growth modes with zero growth rate) that consume the nutrient(s) and produce only non-transporter chemicals. 

In the main text, the results of networks with non-growing subnetworks (I) are mainly presented. However, the results are basically valid even with non-growing subnetworks (II), except that $H$ gets small in $\Sext\to0$ because the components are concentrated on a few free reactants without any reactions occurring to form the intermediate complexes. Notablye, even with a non-growing subnetwork (II), there is a peak of $H$ near the transition point (see Figs.~\ref{fig:RandomNet_DormantII}(c)~and~\ref{fig:RandomNet_Pump_Stat}(b)).

\textit{Death phase.---}
With some reaction networks, there exists a \textit{death phase}, in addition to the growth and dormant phases. Thus, {the dependence on $\Sext$ in our model} can be classified into growth-dormant transition, growth-dormant-death transition, growth-death transition, and no transition. 
The death phase is defined as a stable steady state with zero growth rate with $\Sext>0$ (even though the degradation of internal components is not included in our model). 
When the nutrient concentration is increased from $\Sext=0$ to sufficiently large $\Sext$, cells can return to the growth phase from the dormant phase after a certain lag time, while they cannot from the death phase.

\begin{figure*}[tbh]
\centering 
\includegraphics[width=0.85\linewidth, clip]{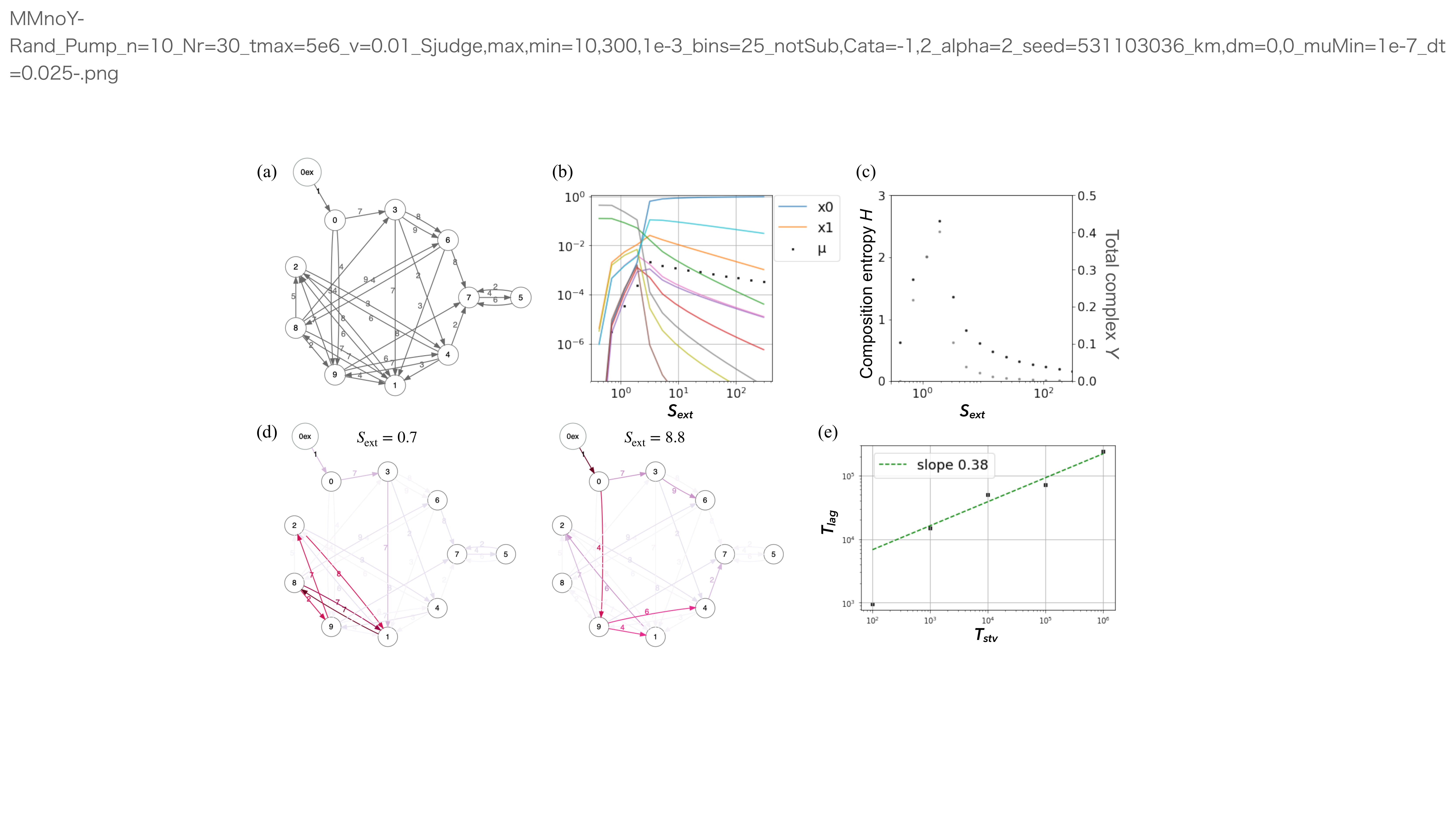}
\caption{
    Example of randomly-generated networks with non-growing subnetwork (II). $n=10,N_r=30$. 
    (a) Reaction network. Chemicals at arrowtails are transformed to those at arrowheads, catalyzed by those labeled on edges. 
    (b) Dependence of $\mu^\ast$ and $\x^\ast$ on $\Sext$. 
    (c) Dependence of $Y := \sum_\rho y_\rho$ and $H := -\sum_ix_i\log x_i-\sum_\rho2y_\rho\log(2y_\rho)$ on $\Sext$. 
    (d) Fluxes for the dormant phase ($\Sext=0.112$; left) and growth phase ($\Sext=10.0$; right). 
	{The edge colors represent the log scale of reaction fluxes.} 
    (e) Dependence of the lag time $\Tlag$ on the starvation time $\Tst$. 
} \label{fig:RandomNet_DormantII}
\end{figure*}

\subsection{Model with degradation}
In the main text, the degradation or leakage of intracellular chemical components is not considered, while one can include it. 
Then, the degradation terms, $-d_ix_i$ and $-d_\rho y_\rho$, should be added to the right-hand side of Eqs.~(\ref{eq:dxdt}-\ref{eq:dydt}) in the main text as:
\begin{eqnarray}
\dot{x}_i &=&
\sum_{\rho} \left[
(\delta_{i,\rho_p}+\delta_{i,\rho_c})v_\rho y_{\rho}-(\delta_{i,\rho_s}+\delta_{i,\rho_c})f_\rho({\bf x},{\bf y})
\right]
 + F_i(\x;\Sext,\alpha) - (\mu+d_i) x_i, \label{eq:dxdt_deg}\\
\dot{y}_\rho &=& f_\rho({\bf x},{\bf y}) - v_\rho y_{\rho} - (\mu+d_\rho) y_\rho,  
\end{eqnarray}
where the growth rate is defined as $\mu(\x,\y) := \sum_iF_i - \sum_id_ix_i - 2\sum_\rho d_\rho y_\rho$. 

The model considered in the main text corresponds to the case with $d_i=d_\rho=0$; whereas, if $d_i,d_\rho>0$, cells reach fixed points with non-positive growth rate, which correspond to death phases, with finite $\Sext$.

\subsection{Forms of nutrient transport}
In the main text, we assumed that the intake of nutrient $X_0$ is mediated by $\alpha$-th order transporter $X_1$, as $F_0(\x;\Sext,\alpha) = \Sext x_{1}^\alpha$. However, other forms of nutrient transport can be considered as well: e.g., $\alpha$-th order channel as $F_i(\x;\Sext,\alpha,D_i)=D_ix_{\sigma_i}^\alpha(\Sext-x_i)$ and passive diffusion without transporters as $F_i(\textbf{x};\Sext,D_i)=D_i(\Sext - x_i)$. 
Even with these forms of nutrient transport, the growth-dormant transition can occur due to a similar mechanism, namely, jamming of reactions due to the accumulation of complexes (see e.g., Fig.~\ref{fig:RandomNet_Channel}). 

\begin{figure*}[tbh]
\centering 
\includegraphics[width=0.85\linewidth, clip]{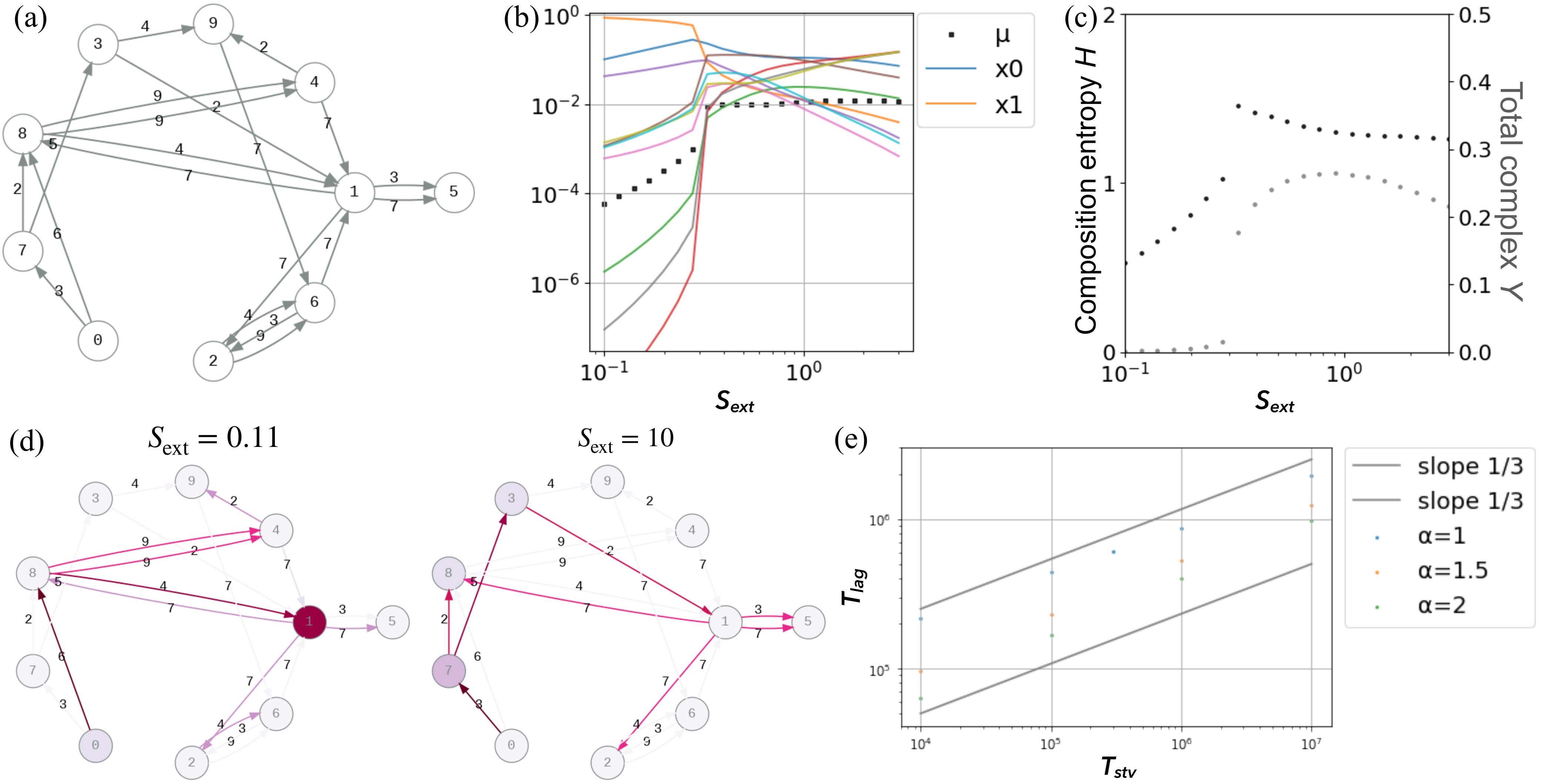}
\caption{
    Example of randomly-generated networks with nutrient transport via the $\alpha$-th order channel $X_1$. $n=10,N_r=20$. Unless otherwise stated, $\alpha=1$ and $v=0.1$.
    (a) Reaction network. Chemicals at arrowtails are transformed to those at arrowheads, catalyzed by those labeled on edges. 
    (b) Dependence of $\mu^\ast$ and $\x^\ast$ on $\Sext$. 
    (c) Dependence of $Y := \sum_\rho y_\rho$ and $H := -\sum_ix_i\log x_i-\sum_\rho2y_\rho\log(2y_\rho)$ on $\Sext$. 
    (d) Dominant fluxes for the dormant phase ($\Sext=0.112$; left) and growth phase ($\Sext=10.0$; right). 
	{The edge colors represent the log scale of reaction fluxes, while the node colors represent the concentrations.} 
    (e) Dependence of the lag time $\Tlag$ on the starvation time $\Tst$ with different $\alpha$. 
} \label{fig:RandomNet_Channel}
\end{figure*}

\subsection{Details about the lag time}
As also described in the main text, {when the intracellular composition in the non-cyclic non-growing subnetwork (II) (working in the dormant or death phase) includes a sufficiently small amount of the transporter chemical, the cell in our model exhibits} the lag time $\Tlag$. In addition,  $\Tlag$ increases with starvation time $\Tst$ in the form $\Tlag\propto\Tst^\beta$ (see e.g., Figs.\ref{fig:RandomNet_DormantII}~and~\ref{fig:RandomNet_Channel}). 

Intuitively, this is understood as follows. Under nutrient scarcity, the chemical concentrations {are gradually concentrated toward the non-growing subnetwork in the dormant phase or death phase; since that absorbing state contains only a small amount or none of some component(s) of the autocatalytic subnetwork of the autocatalytic growth subnetwork, their concentration decreases exponentially.} Then, the cell takes a long time to recover the growth by regaining the intracellular composition in the growth phase after the nutrient supply is recovered. 

The exponent $\beta$, ranging from approximately $0.1$ to $0.5$, seems to be mainly dependent on the network structure which alters the intracellular reaction dynamics: e.g., for the network of Fig.~\ref{fig:RandomNet_Channel}, $\beta\sim 1/3$ holds, almost independently of the exponent $\alpha$ for the nutrient transport.

\section{Details about mean-field analysis}\label{sec:mean-field}
\subsection{Mean-field model with $S$, $X$, and $Y$}
In Fig.~3(a) in the main text, a fully-connected mean-filed model including one mean-filed variable $X$ in addition to the nutrient $S$ is considered. In this model, $X$ represents all $n_X$ non-nutrient chemicals one of which is the transporter; thus, the growth rate is given as $\mu(X/n_X;\Sext) := \Sext(X/n_X)^\alpha$. 

Then, its time evolution is given as 
\begin{eqnarray*} \label{eq:mean-field_simplest}
    \dot{S} &=& \mu(X/n_X;\Sext) - \phi SX - \mu S, \\
    \dot{X} &=& \phi SX - 2\phi (X^2 - vY) - \mu X, \\
    \dot{Y} &=& \phi (X^2 - vY) - \mu Y,
\end{eqnarray*}
where $Y$ denotes the complexes of non-nutrient chemicals and $\phi$ denotes the reaction path density. Here, the complexes between $S$ and $X$ are not explicitly incorporated, in other words, $v_\rho$ for the reactions between $S$ and $X$ are assumed sufficiently large, for the sake of simplicity. 

Even such a mean-field model consisting of only nutrient chemical $S$ and non-nutrient chemical $X$ exhibits the growth-dormant transition with sufficiently small $v \ll \mu_{\max} $ and sufficient non-linearity $\alpha>2$ (Fig.~\ref{fig:mean-field_1var}). 

Even with smaller $\alpha=2$, a transition-like behavior can be observed but it is not a discontinuous transition, as can be seen from the following self-consistent equation for $\mu$ (Fig.~\ref{fig:mean-field_1var}(c)). 
From the steady state condition $\dot{S}=0,\dot{Y}=0$, 
\begin{eqnarray*} \label{eq:mean-field_1variable}
S^\ast = \frac{ \mu }{ \phi_S X + \mu }, \quad
Y^\ast = \frac{\phi_XX^2}{\phi_Xv+\mu}.
\end{eqnarray*}
Then, the steady state $X^\ast$ must satisfy 
\begin{eqnarray*} \label{eq:mean-field_1variable}
    0=\dot{X}= 
    \phi_S \frac{ \mu }{ \phi_S X + \mu }X - 2\phi_X \frac{\mu }{\phi_Xv+\mu}X^2 - \mu X.
\end{eqnarray*}
By solving $X^\ast$ as a function of $\mu$, we obtain
\begin{eqnarray*} \label{eq:mean-field_1variable}
    X^\ast(\mu;\alpha,v,\phi_S,\phi_X) = 
    \frac{-\phi_S \phi_X v - \phi_S \mu -  2 \phi_X \mu + \sqrt{(-\phi_S \phi_X v - \phi_S \mu - 2 \phi_X \mu)^2 +   8 \phi_S \phi_X (\phi_S \phi_X v + \phi_S \mu - \phi_X v \mu - \mu^2)}}
    {4 \phi_S \phi_X}.
\end{eqnarray*}

\begin{figure*}[tbh]
    \centering 
    \includegraphics[width=0.75\linewidth, clip]{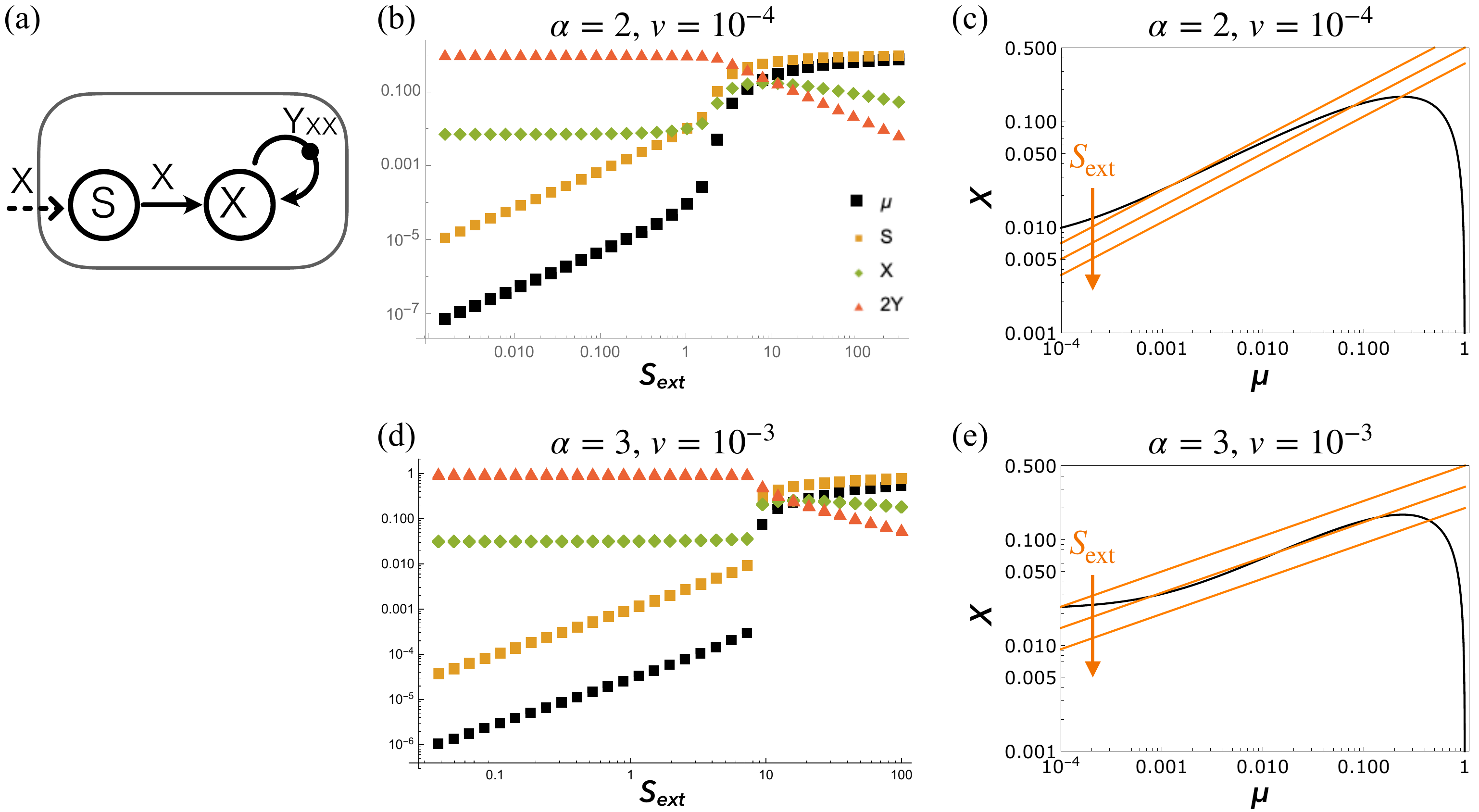}
    \caption{Mean-field model with $S$, $X$, and $Y$. 
    (a) Reaction network structure. 
    (b) Dependence of $\mu^\ast$, $S^\ast$, $X^\ast$, and $Y^\ast$ on $\Sext$. $\alpha=2,v=10^{-4}$. 
    (c) Self-consistent equation for $\mu$ with $\alpha=2,v=10^{-4}$. Orange lines correspond to $\Sext=8,32,128$, respectively.
    (d) Dependence of $\mu^\ast$, $S^\ast$, $X^\ast$, and $Y^\ast$ on $\Sext$. $\alpha=2,v=10^{-4}$. 
    (e) Self-consistent equation for $\mu$ with $\alpha=3,v=10^{-3}$. Orange lines correspond to $\Sext=8,32,128$, respectively.
    } \label{fig:mean-field_1var}
\end{figure*}

\subsection{Simple mean-field model with $S$, $T$, $X$, and $Y$ in the main text}
The dynamics for the simplest ``mean-field'' model with $S$, $T$, $X$, and $Y$ in the main text is given as
\begin{eqnarray*} \label{eq:mean-field_simplest}
    \dot{S} &=& \mu(T;\Sext) - \phi \frac{n_T+n_X}{n_X}SX - \mu S, \\
    \dot{T} &=& \phi\frac{n_T}{n_X}SX - \phi(XT - vY) - \mu T, \\
    \dot{X} &=& \phi SX - \phi(XT - vY) - \mu X, \\
    \dot{Y} &=& \phi(XT - vY) - \mu Y.
\end{eqnarray*}
Note that, in this model, the intermediate complex between $X$ and $T$ is considered for simplicity. Even in such a case, the growth-dormant transition is observed (Fig.~\ref{fig:mean-field_main}(b) in the main text). 

\textit{Self-consistent equation.---}
The self-consistent equation in Fig.~\ref{fig:mean-field_main}(e) in the main text is calculated as follows. 

First, from the definition of growth rate $\mu(\x) := \Sext (T/n_T)^\alpha$, $T = n_T(\mu/\Sext)^{1/\alpha}$ holds. 

In contrast, from the steady state condition, we can explicitly solve $T^\ast$ in the steady state as a function of $\mu$. 
From $\dot{S}=0$ and $\dot{Y}=0$, we immediately obtain $S^\ast(X^\ast,\mu) = \frac{\mu}{\mu + \phi(n_T+n_X)X^\ast/n_X}$ and $Y^\ast(T^\ast,X^\ast,\mu) = \phi\frac{X^\ast T^\ast}{\mu + \phi v}$, respectively. 
Then, by solving $\dot{X}= 0\;\Leftrightarrow\; \phi S^\ast(X^\ast,\mu)X^\ast - \phi(X^\ast T^\ast  - vY^\ast(T^\ast,X^\ast,\mu)) - \mu X^\ast = 0$ in the case $n_T=1,\phi=1,\alpha=2$,
\begin{eqnarray*}  
X^\ast(T^\ast,\mu;v,n_X) = \frac{n_X}{1 + n_X} \frac{v + \mu - v\mu - T^\ast \mu - \mu^2}{v + T^\ast + \mu}.
\end{eqnarray*}
Finally, from $\dot{T}= 0$ and $T^\ast\geq0$, we obtain 
\begin{eqnarray}  \label{eq:mean-field_simplest_T}
T^\ast(\mu;v,n_X) = &&\frac{ -v + n_X v + v^2 + n_X v^2 - \mu + n_X \mu + 4 v \mu + n_X v \mu + 3 \mu^2 }{2 (-v - n_X v - 2 \mu)} 
\nonumber \\
&&-\frac{ (v + \mu) \sqrt{
  1 - 2 n_X + n_X^2 + 2 v + 4 n_X v + 2 n_X^2 v + v^2 + 2 n_X v^2 + 
   n_X^2 v^2 + 2 \mu + 6 n_X \mu + 2 v \mu + 
   2 n_X v \mu + \mu^2}}{2 (-v - n_X v - 2 \mu)}.\nonumber \\
\end{eqnarray}
Notably, this is formally independent of $\Sext$. 

Therefore, the steady states are given as the intersections of $T = \sqrt{\mu/\Sext}$ and $T = T^\ast(\mu;v,n_X)$ (Fig.~\ref{fig:mean-field_main}(e) in the main text).

\begin{figure}[tbh]
    \centering
    \includegraphics[width=0.9\linewidth, clip]{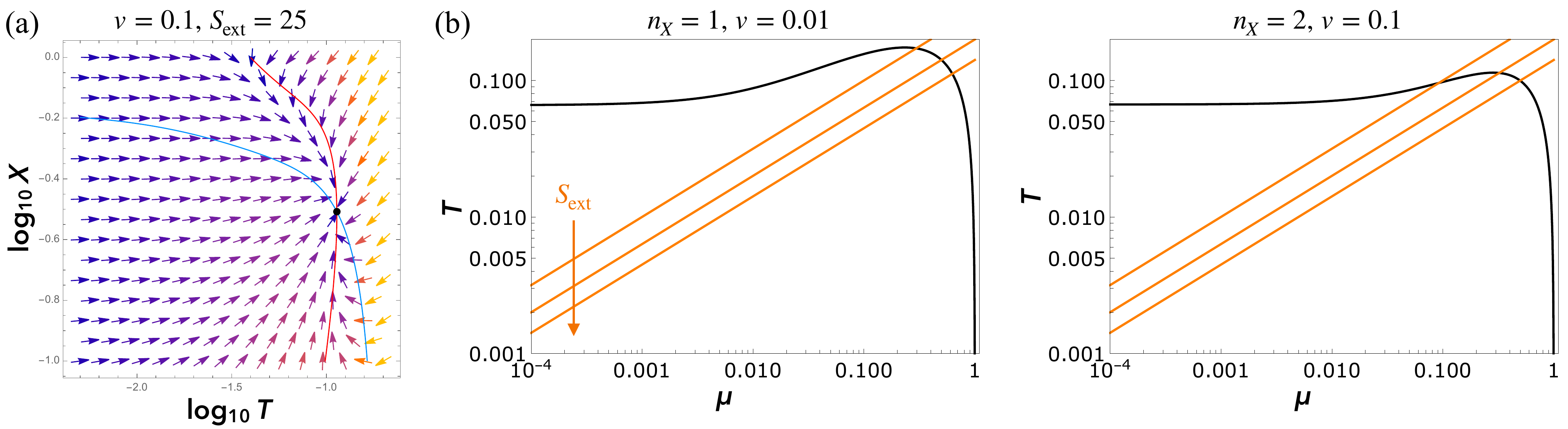}
    \caption{
    {Additional data on the simple mean-field model with $S$, $T$, $X$, and $Y$ (see also Fig.~3(b)-(e) in the main text).} 
    (a) Flow diagram and nullclines with $v=0.1$. 
    (b) Self-consistent equations for $T$ and $\mu$ with $n_X=1,b=0.01$ (left; no transition) and $n_X=2,v=0.1$ (right; no transition). 
    }
    \label{fig:mean-field_simplest}
\end{figure}

\subsection{Fully-connected mean-field model with $S$, $T$, $X$, and their complexes}
We can also consider a fully-connected mean-field model with reactants. Here, all the reactions between $S$, $T$, and $X$ are considered, and the intermediate complex formation (i.e., finite $v_\rho$) for all internal reactions is assumed as in the randomly-generated networks. 
As shown below, this more ``symmetric'' model also exhibits the growth-dormant transition.

The dynamics for the fully-connected mean-filed model is given as \begin{eqnarray*}
\dot{S}&=& \Sext (\frac{T}{n_T})^\alpha - \phi (n_T + n_X) S \frac{X}{n_X} - \mu S,\\
\dot{T}&=& \phi n_T v \frac{Y_{SX}}{n_T + n_X} - \phi n_Tn_X \frac{T}{n_T}\frac{X}{n_X} 
 + \phi n_Tn_X v \frac{Y_{XX}}{(n_T+n_X)n_X}  - \mu T,\\
\dot{X}&=&  - \phi (n_T + n_X) S \frac{X}{n_X} + \phi (n_T + 2n_X) v \frac{Y_{SX}}{n_T + n_X} 
 - \phi n_Tn_X \frac{T}{n_T}\frac{X}{n_X} + 2\phi n_Tn_X v \frac{Y_{TX}}{n_Tn_X} \\
&& - 2\phi (n_T+n_X)n_X \left(\frac{X}{n_X}\right)^2 + \phi (n_T+2n_X)n_X v \frac{Y_{XX}}{(n_T+n_X)n_X} 
 - \mu X,\\
\dot{Y}_{SX}&=& \phi (n_T + n_X) S \frac{X}{n_X} - \phi (n_T + n_X) v \frac{Y_{SX}}{n_T + n_X} -\mu Y_{SX},\\
\dot{Y}_{TX}&=& \phi n_Tn_X \frac{T}{n_T}\frac{X}{n_X} - \phi n_Tn_X v \frac{Y_{TX}}{n_Tn_X} -\mu Y_{TX},\\
\dot{Y}_{XX}&=& \phi (n_T+n_X)n_X \left(\frac{X}{n_X}\right)^2 - \phi (n_T+n_X)n_X v \frac{Y_{XX}}{(n_T+n_X)n_X}  -\mu Y_{XX}.
\end{eqnarray*}
Here, the number of reactions between chemicals $i$ and $j$ are proportional to $n_in_j$ ($i,j=S,T,X$). 
When $Y_\rho$s are adiabatically eliminated as 
\begin{eqnarray*}
Y_{SX}^\ast = \frac{\phi(n_T+n_X)/n_X}{\phi v + \mu} SX
,\quad Y_{TX}^\ast = \frac{\phi }{\phi v + \mu} TX
,\quad Y_{XX}^\ast = \frac{\phi (n_T+n_X)/n_X}{\phi v + \mu} X^2,
\end{eqnarray*}
we obtain: 
\begin{eqnarray*}
\dot{S} &=& \Sext (\frac{T}{n_T})^\alpha - \phi (n_T + n_X) S \frac{X}{n_X} - \mu S,\\
\dot{T} &=& \phi_S n_T v \frac{\phi_S/n_X}{\phi_Sv + \mu} SX - \phi TX 
 + \phi n_T v \frac{\phi /n_X}{\phi v + \mu} X^2  - \mu T,\\
\dot{X} &=&  \frac{n_X\phi_S v -(n_T + n_X)\mu }{\phi_Sv + \mu}\phi_S S\frac{X}{n_X}
 + \frac{\phi v - \mu}{\phi v + \mu}\phi TX 
 -\phi\frac{2 (n_T+n_X) \mu + n_T v \phi}{\phi v + \mu}X^2 /n_X  - \mu X.
\end{eqnarray*}

This fully-connected mean-field model exhibits the growth-dormant transition, (Fig.~\ref{fig:mean-field_2var_FullConnected}). 

\begin{figure*}[tbh]
    \centering 
    \includegraphics[width=0.5\linewidth, clip]{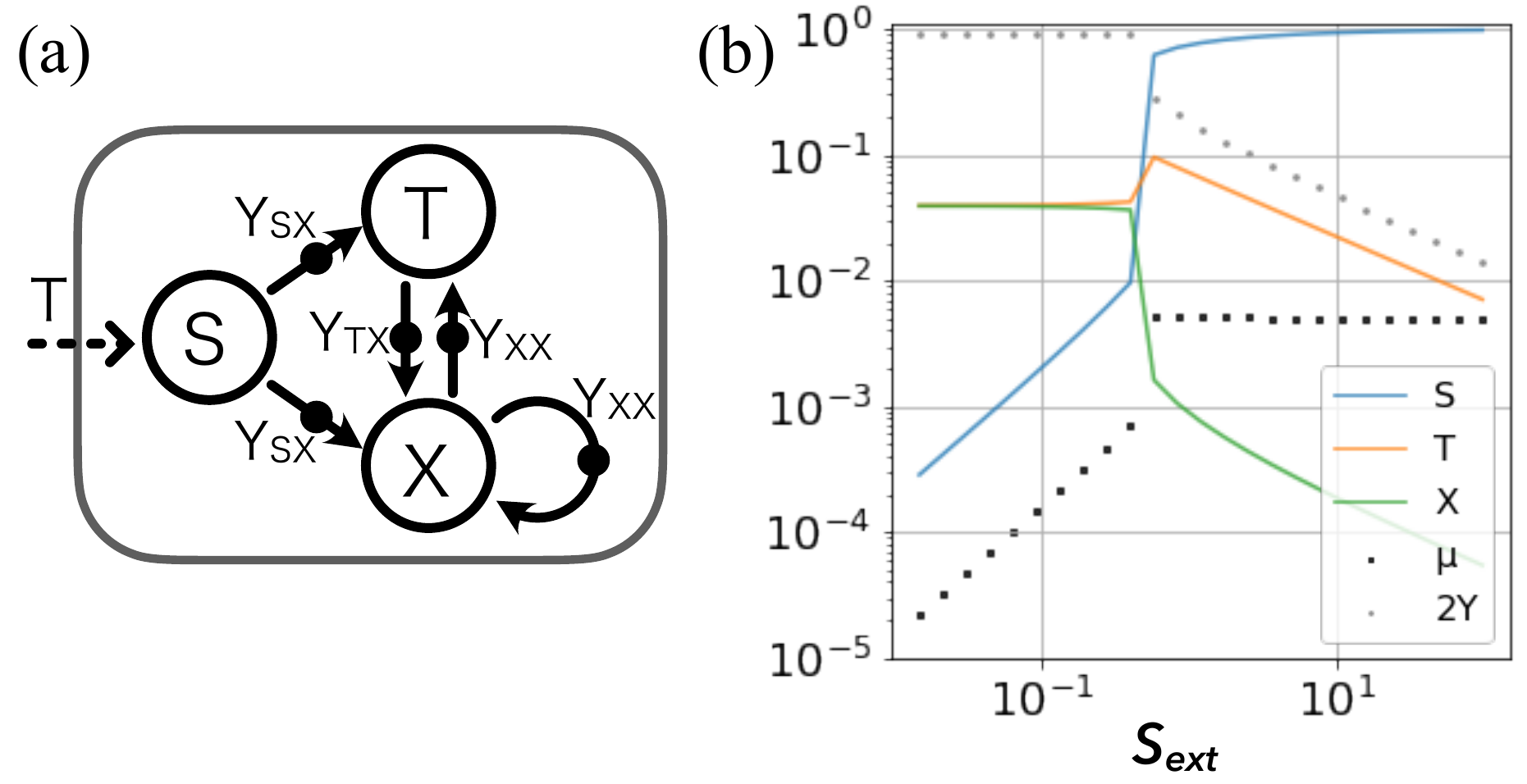}
    \caption{Fully-connected mean-field model with $S$, $T$, $X$, and their complexes. 
    (a) Reaction network structure. 
    (b) Dependence of $\mu^\ast$, $S^\ast$, $X^\ast$, $T^\ast$, and $Y:=Y^\ast_{SX}+Y^\ast_{XX}+Y^\ast_{TX}$ on $\Sext$. $v=0.01$. 
    } \label{fig:mean-field_2var_FullConnected}
\end{figure*}

\end{document}